\documentclass[12pt]{article}

\usepackage{url} 
\usepackage{amsmath,amsthm,amssymb}
\usepackage{graphicx,psfrag,epsf}
\usepackage{enumerate}
\usepackage{bm}
\usepackage{enumitem}
\usepackage{multirow}
\usepackage{scalerel}
\usepackage{float}
\usepackage[authoryear,round]{natbib}
\usepackage{adjustbox}
\usepackage[toc,page]{appendix}
\usepackage[dvipsnames]{xcolor}
\usepackage{booktabs}
\usepackage{comment}

\usepackage[linesnumbered,ruled,vlined]{algorithm2e}

\SetKwInput{KwInput}{Input}                
\SetKwInput{KwOutput}{Output}              

\def\squarebox#1{\hbox to #1{\hfill\vbox to #1{\vfill}}}
\def\boxit#1{\vbox{\hrule\hbox{\vrule\kern6pt
          \vbox{\kern6pt#1\kern6pt}\kern6pt\vrule}\hrule}}

\theoremstyle{definition}

\newtheorem{lemma}{Lemma}
\newtheorem{thm}{Theorem}
\newtheorem{pro}{Proposition}
\newtheorem{remark}{Remark}

\newcommand{\bh}{\mathbf{h}}

\newcommand{\hs}{h^{\star}}

\newcommand{\setI}{\mathbb{I}}
\newcommand{\setR}{\mathbb{R}}
\begin{document}
\begin{center}
	{\Large{\bf Efficient Estimation for Functional Accelerated Failure Time
Model}}
		
\end{center}
	\begin{center}	
		{\large Changyu Liu$^{1}$, Wen Su$^{2}$, Kin-Yat Liu$^1$, Guosheng Yin$^2$ and Xingqiu Zhao$^{1}$}\\ 
$^1${Department of Applied Mathematics,
			The Hong Kong Polytechnic University,
			Hong Kong\\cy.u.liu@connect.polyu.hk; kin-yat.liu@connect.polyu.hk; xingqiu.zhao@polyu.edu.hk\\
$^2${Department of Statistics and Actuarial Science, University of Hong Kong, Hong Kong\\ jenna.wen.su@connect.hku.hk; gyin@hku.hk }		
}
	\end{center}

	\begin{abstract}
	We propose a functional accelerated failure time  model to  characterize effects of both functional and scalar covariates on
the time to event of interest, and provide regularity conditions to guarantee  model identifiability.
For efficient estimation of model parameters, we develop a  sieve maximum likelihood approach where parametric and nonparametric coefficients are bundled with an unknown baseline hazard function in the likelihood function.
Not only do the bundled parameters cause immense numerical difficulties, but they also
result in new challenges in theoretical development. By developing a general theoretical framework,
we overcome the challenges arising from the bundled parameters and derive the convergence rate of the proposed estimator.
Furthermore, we prove that the finite-dimensional estimator
is $\sqrt{n}$-consistent, asymptotically normal and achieves the semiparametric information bound. The proposed inference procedures are evaluated by extensive simulation studies and illustrated with  an application  to
the sequential organ failure assessment data from the Improving Care of Acute Lung Injury Patients study.
   \end{abstract}
    \noindent%
    \vfill

    \newpage
    \section{Introduction}
Functional data  are typically  regarded as a realization of an underlying stochastic process; that is,
$Z(\cdot): \setI_0\rightarrow \mathbb{R}$ is a stochastic process indexed with a compact set $\mathbb{I}_0$.
Technological advancement has drastically increased the capability of capturing and storing
functional data, which became increasingly important in many fields such as medicine, economics, engineering
and chemometrics. With growing awareness of its importance,  a vast amount of  literature
has been devoted to the development of functional data analysis, of which functional regression analysis
has received the most attention in application and methodology development (Morris, 2015).
The functional linear model (FLM) was first introduced by \cite{Ramsay1991}, which was later
extended to various nonlinear functional models,
including  the generalized functional linear model (Marx and Eilers, 1999), 
the functional polynomial model (Yao and M\"{u}ller, 2010), 
and the functional generalized additive model (McLean et al., 2014). 
For the problem of prediction and estimation,  the approaches based
on the functional principal component analysis (FPCA) have been commonly used
(Cardot, Ferraty, and Sarda, 1999; M\"{u}ller and Stadtm\"{u}ller, 2005; Yao, M\"{u}ller, and Wang, 2005; Crainiceanu, Staicu, and Din, 2009).
Morever, other methods including different basis functions and regularization approaches
have also been well developed. Additional details and insights of functional data analysis are
discussed in the monographs   by \cite{Ramsay2005} and  \cite{Ferraty2006} as well as the
reviews by \cite{Morris2015} and \cite{Wang2016}.

Recently, functional data received substantial amount of attention in the realm of survival analysis.
\citet{Chen2011} proposed the functional Cox model and \cite{Kong2018} extended this model to the FPCA approach.
\cite{Qu2016} studied the model estimation under
a more general reproducing kernel Hilbert space framework, where they derived
the asymptotic properties of the maximum partial likelihood estimator and established the asymptotic normality and efficiency for the  finite-dimensional estimator. Furthermore,   \cite{Hao2020}  derived  the asymptotic joint distribution of finite-  and infinite-dimensional estimators. \cite{Cui2021} proposed the additive functional Cox model.
Jiang et al. (2020) studied a functional censored quantile regression model to
characterize the time-varying relationship
between time-to-event outcomes and corresponding functional covariates.
Yang et al. (2020) considered the functional linear regression model for right-censored data and
developed a penalized least squares method for model estimation, while the theoretical properties
of the proposed estimator have not been studied yet.

Among various survival models, the Cox proportional hazards model (Cox, 1972) has gained
the most popularity in applications. However, when the proportional hazards assumption is violated
as encountered commonly in practice,
the accelerated failure time (AFT) model provides a convenient and attractive alternative
to regression analysis for censored data (Buckley and James, 1979; Miller and Halpern, 1982; Ritov, 1990;
Tstats, 1990; Lai and Ying, 1991a, 1991b; Ying, 1993; Jin et al., 2003; Jin, Lin, and Ying, 2006;
Zeng and Lin, 2007; Ding and Nan, 2011; Lin and Chen, 2013). With the transformed failure time directly
regressed on the covariates, the AFT model has the advantage of straightforward interpretation inherited
from the typical linear regression.
To accommodate for both functional and scalar data, we consider a functional accelerated failure time (FAFT) model,
\begin{equation}\label{model}
T=\alpha_{0}^{\top}X+\int_{\setI_0}\beta_{0}(s)Z(s)ds+\varepsilon,
\end{equation}
where   $T$ is  a  failure time after a known monotone transformation, $X$ is  a $p$-dimensional vector of covariates,  $Z(\cdot)$  is a functional covariate, $\alpha_{0}$ is a $p$-dimensional parameter, $\beta_{0}(\cdot)$ is a functional parameter, and $\varepsilon$ is an error with an unknown distribution.

In this paper, we develop a sieve maximum likelihood approach for the  FAFT model   with  right-censored data.
To investigate the asymptotic property of the sieve estimator,  we need to overcome   two main challenges.  First, the parameters are bundled together in the log-likelihood function such that the theoretical analysis is much
more difficult than the usual situations with separate parameters in objective functions.
Second, the overall convergence rate of the proposed estimator is shown to be lower than the standard rate
$n^{-1/2}$, which brings considerable difficulties in deriving the asymptotic distribution of the estimator.

The main contributions are as follows:
\begin{itemize}
	\item[(i)] We rigorously discuss the model identifiability and provide sufficient conditions. Even for the AFT model with an unspecified error distribution, the existing statistical inference procedures are typically made by assuming the model to be identifiable.
	\item[(ii)] Overcoming the challenges from bundled parameters, we establish the convergence rate of
the bundled parameter (Theorem 1) and those of other parameters (Theorem 2). Our theoretical development is highly nontrivial
and general enough to be applicable to other bundled parameter situations.
	\item[(iii)] We obtain the information bound for the finite-dimensional parameters in the semiparametric FAFT model
and demonstrate the efficiency of our estimation procedure.
	\item[(iv)] We derive the asymptotic normality for  the finite-dimensional estimator   and show that it achieves information bound asymptotically.   Therefore,  the proposed estimation approach is asymptotically efficient.
	\end{itemize}

The rest of this paper is organized as follows. In Section \ref{sec:pre},
we present  the log-likelihood function  and the estimation procedure. In Section \ref{sec:result}, we state the regularity conditions and summarize the asymptotic properties of the proposed estimators. Simulation studies and applications to the sequential organ failure assessment data are described in Sections \ref{sec:numerical studies} and \ref{sec:application}, respectively. Some concluding remarks are made in Section \ref{sec:conclusion}. Proof of theorems and technical details are relegated to the supplementary materials.

\section{Estimation Method}\label{sec:pre}

Let  $U=\left(X, Z(\cdot)\right)$ denote the covariate and $\theta=(\alpha,\beta(\cdot))$  denote the   parameter. Define $\mu(U,\theta)=\alpha^{\top}X+\int_{\setI_0}\beta(s)Z(s)ds$.
The FAFT model in \eqref{model} can be rewritten as
\begin{equation}\label{model1}
	T=\mu(U,\theta_{0})+\varepsilon,
\end{equation}
where $\theta_{0}=(\alpha_{0},\beta_{0}(\cdot))$ is the true parameter. Let $C$ denote the  censoring time after the same transformation as the failure time $T$.  The observed survival time is $Y=\min\{T, C\}$ with censoring  indicator $\Delta=I(T\leq C)$. Under a standard assumption that $\varepsilon$ is independent of $U$ and $C$, subsequently we have $T$ and $C$ being independent conditional on covariate $U$. Hence the joint density function of $(Y, \Delta, U)$  is
\begin{equation*}
	f_{Y,\Delta,U}(y,\delta,u) =\lambda_{0}^{\delta}\big(y-\mu(u,\theta_0)\big)\exp\big\{-\Lambda_0\left(y-\mu(u,\theta_0)\right)\big\}H(y,\delta,u),
\end{equation*}
where $\lambda_{0}(\cdot)$ and $\Lambda_0(\cdot)$ are the hazard function and the cumulative hazard function of the error term $\varepsilon$, respectively,  and $H(y,\delta,u)$ is a function that  depends only on the distribution of $U$ and the conditional distribution of $C$ given $U$. In order to alleviate the positivity  constraint for the hazard function, we set  $g(\cdot)=\log\lambda(\cdot)$ and formulate the log-likelihood function as a function of $(\theta, g)$.

Suppose the observations  $\left(Y_{i}, \Delta_{i}, U_{i}\right),$ $ i=1,\dots, n, $ are independently sampled based on the FAFT model,
and then the log-likelihood function for the  parameter $\xi=(\alpha,\beta, g)$ is given by
\begin{equation}\label{log-mle}
l_n(\xi)=\frac{1}{n}\sum_{i=1}^{n}\bigg\{\Delta_{i} g\big(Y_i-\mu(U_i,\theta)\big)-\int \exp\{g(t)\}I\left\{  Y_i-\mu(U_i,\theta) \geq t\right\}dt\bigg\},
\end{equation}
where the parts independent of $\xi$ are omitted.  We consider estimating $\xi$ by maximizing  the log-likelihood function,
for which direct estimation is infeasible. \cite{Zeng2007} showed that the maximum of $\ell_n(\xi)$ does not exist even when all
the covariates are scalar (i.e., no functional component).  To overcome this difficulty, \cite{Zeng2007} proposed a kernel-smoothed profile likelihood function for the estimation of  regression parameters.   \cite{Nan2011} investigated the model by applying the spline method. However, none of these methods can be applied to the FAFT model due to the inclusion of the
functional component which not only causes numerical challenges but also theoretical difficulties.


We propose an estimation approach for the FAFT model by maximizing the log-likelihood function in a sieve space.  Specifically, we  focus on the spline-based sieve space,  where both scalar and functional parameters are estimated simultaneously as bundled together.
The advantages of this spline-based sieve space are demonstrated both theoretically and numerically. The choice of sieve space is general  as long as the assumptions for the theorems are satisfied.

Without loss of generality, we assume $\setI_0=[0,1]$ and the log-hazard function $g_{0}$ is supported on $[a,b]$, as an interval of interest, where $a=\inf_{y,u}\{y-\mu(u,\theta_{0})\}$ and $b=\tau<\infty$.    To propose the  spline-based sieve space,  we  first introduce  some   notation. For a  closed interval $[c,d]$, let $\mathcal{T}_{n}(c,d)=\{t_{i}, i=0,\dots, m_{n}+1\}$ denote a sequence of knots that partition $[c,d]$ into $m_{n}+1$ subintervals, where  $c\equiv t_{0}<t_{1}<\dots<t_{m_{n}}<t_{m_{n}+1}\equiv d.$
Let  $\mathcal{S}_{\ell}\{\mathcal{T}_{n}(c, d)\}$ denote the space of splines of order $\ell\geq 1$ with the knot sequence $\mathcal{T}_{n}(c, d)$ and let $q_{n}= m_n+\ell$.  According to Corollary 4.10 of \cite{Schumaker1981},  for any function $\phi\in \mathcal{S}_{\ell}\{\mathcal{T}_{n}(c, d)\}$, there exists a $q_{n}$-dimensional vector $\gamma$ such that
  $\phi =B_{n}^{\top}\gamma$, where $B_{n}=(b_1,\dots, b_{q_{n}})^{\top}$ is a vector of B-spline  basis functions.
  Following \cite{Shen1994},  we consider the space
  \begin{equation*}
   \Phi_{n}\big(\ell,c,d\big) =\Big\{B_{n}^{\top}\gamma: \;\|\gamma\|_{\infty}\leq c_{n}\Big\},
  \end{equation*}
  where $c_{n}$  grows with $n$ slowly enough.
 Define  $\mathcal{F}^{\omega}_{n}=\Phi_{n}(\lceil \omega\rceil+1,0,1)$ and $\mathcal{G}_{n}^{\kappa}=\Phi_{n}(\lceil \kappa\rceil+1, a, b)$, where $\lceil x\rceil$ is the ceiling function,   $\omega$ and $\kappa$
 respectively represent the smoothness  of $\beta$ and $g$ in Condition \ref{assume4} given in the next section.
The sieve space is defined as $$\Xi_{n}=\mathcal{B}\times \mathcal{F}^{\omega}_{n}\times \mathcal{G}_{n}^{\kappa}=\Big\{
  \xi=\big(\alpha, \beta, g\big): \alpha\in\mathcal{B}, \beta\in   \mathcal{F}^{\omega}_{n}, g\in  \mathcal{G}_{n}^{\kappa}
  \Big\},$$
 where $\mathcal{B}$ is a known compact set of $\mathbb{R}^{p}$. We study the following   sieve maximum likelihood estimator:
  \begin{equation}\label{sieve-mle}
  	\hat{\xi}_{n}=\big(\hat{\alpha}_{n}, \hat{\beta}_{n}, \hat{g}_{n}\big):=\underset{\xi\in\Xi_{n}}{\arg\max}\ l_n(\xi).
  \end{equation}
Under Condition \ref{assume5} in the next section, it is equivalent to find a
$(p+m_{n}^{\omega}+s_{n}^{\kappa})$-dimensional vector $( \alpha^{\top}, \gamma^{\beta \top}, \gamma^{g \top})^{\top}$
that maximizes the log-likelihood function by taking $\beta=\gamma^{\beta \top}B_n^{\beta}$
and $g=\gamma^{g \top}B_n^{g}$, where $B_n^{\beta}$ and $B_n^{g}$ are the vectors  of B-spline basis functions of $\mathcal{F}_n^{\omega}$ and $\mathcal{G}_n^{\kappa}$, respectively. Therefore, for the sieve maximum likelihood estimator $	\hat{\xi}_{n}$, there exists an $m_n^{\omega}$-dimensional vector $\hat{\gamma}_{n}^{\beta}$ and an $s_n^{\kappa}$-dimensional vector $\hat{\gamma}_{n}^{g}$ such that $\hat{\beta}_{n}=\hat{\gamma}_{n}^{\beta \top}B_n^{\beta}$ and $\hat{g}_{n}=\hat{\gamma}_{n}^{g \top}B_n^{g}$.

The estimate of $( \alpha^{\top}, \gamma^{\beta \top}, \gamma^{g \top})^{\top}$ is then obtained by maximizing the following log-likelihood function,
\begin{equation*}
l_n(\xi)=\frac{1}{n}\sum_{i=1}^{n}\bigg\{\Delta_{i} \gamma_{n}^{g \top} B^g_n\big(Y_i-\mu( U_i,\theta)\big)-\int e^{\gamma_{n}^{g \top} B^g_n(t)}I\left\{  Y_i-\mu(U_i,\theta) \geq t\right\}dt\bigg\},
\end{equation*}
where $\mu(U_i,\theta)=\alpha^{\top}X_i+\int_{\setI_0}\gamma_{n}^{\beta \top} B^\beta_n(s)Z_i(s)ds$.

The aforementioned estimation can be easily implemented in R software. The integration was carried out by the \texttt{integrate} function, the B-spline basis functions were constructed by the \texttt{create.bspline.basis} function from the \texttt{fda} R package (Ramsay, Graves, and Hooker, 2020), and maximization was performed by the \texttt{optim} function. For the maximization procedure, quasi-Newton methods were preferred over Newton's method to mitigate lengthy computational time required to calculate
the Hessian matrix in each iteration. Algorithm 1 outlines the estimation procedure.

\begin{algorithm}[!ht]
\DontPrintSemicolon
  \KwInput{Initial value $\big(\alpha_0, \gamma^\beta_0, \gamma^g_0\big)$}
  \KwOutput{Maximizer $\big(\hat{\alpha}_{n}, \hat{\gamma}^\beta_{n}, \hat{\gamma}^g_{n}\big)$}
  \KwData{Input $(Y_i,\Delta_i,U_i),i=1,\ldots,n$}
  \begin{description}
  \item[Step 1.] Create B-Spline basis functions, $B^\beta_{n,i}(\cdot)$ and $B^g_{n,j}$,
  where $i=1,...,m^\omega_n$ and $j=1,...,s^\kappa_n$.

 \item[Step 2.] Compute $\l_n(\cdot)$ and $\partial l_n(\cdot)/\partial \xi$.

  \item[Step 3.] Obtain the maximizer of $l_n(\cdot)$ using a quasi-Newton method.
\end{description}
\caption{The Algorithm of Estimating $( \alpha, \gamma^{\beta}, \gamma^g)$ }
\end{algorithm}


\section{Theoretical Results}\label{sec:result}
Let  $r_1$ be a positive integer and $r_2\in(0,1]$ such that $r=r_1+r_2$.  Define  $\mathcal{F}_{r}(\mathbb{I})$ as
the class of functions $f$ on $\setI$ whose $r_1$-th derivative exists and satisfies  a Lipschitz condition of order $r_2$; that is,
\begin{equation*}
	\begin{split}
		\mathcal{F}_{r}(\mathbb{I})=\Bigl\{ f:\mathbb{I}\rightarrow \mathbb{R} \mid&\; f \text{  has bounded derivatives } f^{(j)}, j=1,\dots, r_1, \\&\text{and }
		|f^{(r_1)}(s)-f^{(r_1)}(t)|\leq L|s-t|^{r_2} \; \text{for }  s,t\in  \mathbb{I}\Bigr\},
	\end{split}
\end{equation*}
where  $L$ is a positive constant.  Define $r_{\theta}=Y-\mu(U,\theta)$ and $r_{\theta_{0}}=Y-\mu(U,\theta_{0})$.   To establish the asymptotic properties of the proposed estimator, we need the following conditions:
\begin{enumerate}[label=(A\arabic*)]
\item \label{assume1}
The true parameter $\alpha_{0}$ belongs to the interior of a compact set $\mathcal{B}\subseteq\mathbb{R}^{p}.$

\item\label{assume2}
(i) The  covariate $X$ takes values   in a bounded subset $\mathcal{X}\subseteq\mathbb{R}^{p}$    and   satisfies  $E(X)=0$ and $E(XX^{\top})$ is nonsingular; (ii) The $L_2$-norm  of $Z$ is bounded almost surely and  $E(Z)=0$.

\item \label{assume3}
There is a truncation time $\tau<\infty$ such that, for some constant $\delta$, $P\left(r_{\theta_{0}}>\tau\mid U\right)\geq \delta>0$ almost surely with respect to the probability measure of $U$. This implies that $\Lambda_0(\tau)\leq -\log \delta<\infty.$

\item \label{assume4}
The true functional parameter $\beta_{0}$ belongs to $\mathcal{F}^{\omega}\equiv \mathcal{F}_{\omega}\left([0,1]\right)$, where $\omega\geq 1$. The true log-hazard function $g_{0}$ belongs to $\mathcal{G}^{\kappa}\equiv \mathcal{F}_{\kappa}\left([a,b]\right)$, where $\kappa\geq 3$, and $g_0$ is a non-constant and non-periodic function.

\item \label{assume5}
\begin{enumerate}
	\item [(i)]	For   $\mathcal{F}_{n}^{\omega}$,  let  $\mathcal{T}_{n}(0,1)=\{t_{i}, i=0,\dots, m_{n}+1\}$ denote the corresponding knot sequence.   The maximum spacing of  the knots satisfies $  \max_{1\leq i\leq m_{n}+1 } |t_i-t_{i-1}|=O(n^{-\nu})$ and $m_{n}=O(n^{\nu})$ for $\nu\in(0,0.5)$. Define $m^{\omega}_{n}=m_{n}+\lceil\omega\rceil+1.$
	\item [(ii)]For  $\mathcal{G}^{\kappa}_{n}$,  let  $\mathcal{T}_{n}(a,b)=\{t_{i}, i=0,\dots, s_{n}+1\}$ denote the corresponding knot sequence. The maximum spacing of the knots satisfies $ \max_{1\leq i\leq s_{n}+1} |t_i-t_{i-1}|=O(n^{-q})$ and $s_{n}=O(n^{q})$ for $q\in(0,0.5)$.  Define $s^{\kappa}_{n}=s_{n}+\lceil\kappa\rceil+1.$
\end{enumerate}
\item \label{assume6}
For some $\eta\in(0,1)$, $Var(\mu(U,\theta)|r_{\theta_0})\geq \eta E(\mu(U,\theta)^{2}|r_{\theta_0})$ holds almost surely for  any $\theta\in\mathcal{B}\times\mathcal{F}^{\omega}.$

\item \label{assume7}
 The $(\kappa-1)$-th partial derivative of the joint density $f(t,x,z,\Delta=1)$ of $(\varepsilon, X,Z,\Delta=1)$ with respect to $t$ exists and is bounded.
\end{enumerate}
Conditions \ref{assume1}   and \ref{assume4}  place restrictions on the parameter space, which require $\alpha_{0}$ is not on the boundary of the parameter space as well as $g_{0}$ and $\beta_{0}$ satisfying certain smoothness. Such smoothness assumption is often
adopted in nonparametric  estimation and can be easily satisfied.  Similar  regularity conditions are commonly imposed in the  literature [see \cite{Huang1999}, \cite{Zeng2007}, and \cite{Nan2011}].  Condition \ref{assume5} is a regularity condition about the spline-based sieve space.  Condition \ref{assume2}   places a boundedness restriction on the covariates, which is also assumed by \cite{Qu2016}. Condition \ref{assume3} is the same as that in \cite{Nan2011}. Condition  \ref{assume6} guarantees that the convergence rate of each parameter can be derived from the result of bundled parameter $g(r_{\theta})$. Condition \ref{assume7} is required for showing that the score functions of the functional parameters in the least favorable direction  are nearly zero, which is a key step in the derivation of the asymptotic normality of the scalar estimator.

 Define the parameter space as
 \[
 \Xi=\mathcal{B}\times\mathcal{F}^{\omega} \times \mathcal{G}^{\kappa}=\Big\{\xi=\big(\alpha, \beta, g\big): \alpha\in\mathcal{B}, \beta\in   \mathcal{F}^{\omega} , g\in  \mathcal{G}^{\kappa}
 \Big\}.
 \]
Intuitively, $\Xi_{n}\subseteq\Xi_{n+1}\subseteq\cdots\subseteq \Xi$ for all $n\geq 1$.   For notational simplicity,  we also denote $\xi=(\alpha,\beta, g)$   by $\xi=(\theta, g)$ with $\theta=(\alpha, \beta)$. For the parameter space $ \Xi$, we define  the metric $d(\cdot,\cdot)$ as
 \begin{equation*}
 d\big(\xi_1,\xi_2\big)^2=P\big(\mu\left(U,\theta_1-\theta_2\right)^2\big)+\big\|g_1-g_2\big\|_{\mathcal{G}}^2,
 \end{equation*}
 where $P$ denotes the  probability measure with $Pf=\int fdP$, and   $\|g\|_{\mathcal{G}}^2  = P\left(   \Delta  g(r_{\theta_{0}})^2\right).$
 Based on this  metric,   the accuracy of $\hat{\alpha}_n$ can be  measured by the usual  Euclidean   norm $|\hat{\alpha}_n-\alpha_0|$, and the accuracy of $\hat{\beta}_n$ can be measured by the weighted $L_{2}$-norm $\|\hat{\beta}_n-\beta_{0}\|_{C}$, where
 \[
 \big\|\beta\big\|_{C}^{2}=\iint \beta(s) C(s, t) \beta(t) ds dt \; \text{  with  } \; C\big(s, t\big)=\operatorname{E}\big( Z(s) Z(t)\big).
 \]
 The norm $\|\cdot \|_{C}$ has
 been widely used for functional linear models (Cai and Yuan, 2012) 
 and has also been investigated in the functional Cox model  (Qu, Wang, and Wang, 2016).

Direct investigation of the estimator under the metric $d(\cdot,\cdot)$ is challenging because the parameters $g$ and $\theta$ are  bundled together, which makes the information of separate parameters difficult to derive. To overcome this difficulty,  we propose to first investigate the space for bundled parameters, and then apply the results to study parameters separately.   To this end,    we define the  space   of bundled  parameters as
  \begin{equation*}
  	\begin{split}
  	\mathcal{A} = \Big\{g\big(y-\mu(u,\theta)\big):  \xi =(\theta, g)\in\Xi\Big\}.
  	\end{split}
  \end{equation*}
 {For any given $\xi=(\theta, g)$,  the element in $\mathcal{A}$ is denoted  by $g(r_{\theta})$ when  there is no confusion.} To  measure the difference between any two elements in $\mathcal{A}$, we consider the pseudometric:
   $$
   \big\| g_1(r_{\theta_1})-g_2(r_{\theta_2})\big\|_{\mathcal{A}} = P\Big(\Delta \big\{g_1(r_{\theta_1})-g_2(r_{\theta_2})\big\}^2\Big)^{1/2}.
   $$	
  We  first derive the efficient  score function and the information bound.
  \begin{pro}\label{pro:score}
Under Conditions \ref{assume1}--\ref{assume4} and \ref{assume6},  	the efficient score function for estimating $\alpha_0$ in the FAFT model   is
  	\begin{equation*}
  	\dot{l}_{\alpha_0}^{\star}=\int \left\{-\dot{g}_0(t) X + \dot{g}_0(t) \int_0^1 b^{\star}(s)Z(s)ds- \phi^{\star}(t)\right\} dM(t),
  	\end{equation*}
  	where $M(t)=\Delta I(r_{\theta_{0}}\leq t)-\int^{t}_{-\infty}I(r_{\theta_{0}}\geq u)\lambda_{0}(u)du$ and $(b^{\star}, \phi^{\star})$  is a solution that minimizes
  	\begin{equation*}
  	E\left[\Delta \Big|-\dot{g}_0(r_{\theta_0}) X + \dot{g}_0(r_{\theta_0}) \int_0^1 b(s)Z(s)ds- \phi( r_{\theta_0})\Big |^2\right].	
  	\end{equation*}
  	The information bound for estimation of $\alpha_0$ is
  	\begin{equation*}
  	I\left(\alpha_0\right)=E\left [\dot{l}^{\star\otimes2}_{\alpha_0}\right]=E\left[\Delta \big(-\dot{g}_0(r_{\theta_0}) X + \dot{g}_0(r_{\theta_0}) \int_0^1 b^{\star}(s)Z(s)ds- \phi^{\star}( r_{\theta_0})\big)^{\otimes2}\right],
  	\end{equation*}
  	where $x^{\otimes 2}=xx^{\top}$ for any vector $x\in\mathbb{R}^{p}.$
  \end{pro}
 \begin{pro}\label{pro:identifiability}
	For any $\xi^{\star}=(\theta^{\star}, g^{\star})$ that maximizes $E l_n(\xi)$, it satisfies $\|g^{\star}(r_{\theta^{\star}})-g_{0}(r_{\theta_0})\|_{\mathcal{A}}=0.$  Under Conditions \ref{assume1}--\ref{assume4}, we further have $d(\xi^{\star}, \xi_{0})
	=0$. Moreover,    if  $I(\alpha_0)$ is nonsingular, then $g^{\star}=g_{0}$, $\alpha^{\star}=\alpha_0$, and $\beta^{\star}=\beta_0$.
\end{pro}
 \begin{remark}
 The result of Proposition \ref{pro:identifiability} provides sufficient conditions to guarantee identifiability of model (1).
Such identifiability is the key to statistical inference. However,  in the AFT model, accelerated hazards regression model, and longitudinal data model, statistical inference is often based on a direct assumption of model identifiability (e.g., Zeng and Lin, 2007; Zhao, Wu, and Yin, 2017; Kong et al., 2018).
\end{remark}

We next give the convergence rate of the  bundled estimator $\hat{g}_n(r_{\hat{\theta}_{n}})$.
\begin{thm}\label{thm2}
Under Conditions \ref{assume1}--\ref{assume6}, we have
$$ \big\|\hat{g}_n(r_{\hat{\theta}_{n}})-g_{0}(r_{\theta_{0}})\big\|_{\mathcal{A}}=O_p(n^{-c}),$$
 where $c=\min\{\omega\nu, \kappa q, (1-\max\{\nu,q\})/2\}.$
\end{thm}


Here, the consistency of each estimator is derived separately. For $\hat{g}_n$, we first show that the sequence  $\{\hat{g}_n\}_{n\geq 1}$ is precompact and then apply the Arzel$\grave{a}$-Ascoli theorem. The result indicates  that both $\hat{g}_n$ and $\dot{\hat{g}}_n$   converge  in probability under the supremum norm $\|\cdot\|_{\infty}$. Similar approaches were studied in  \cite{Wellner1999} and \cite{Kuchibhotla2020}, however rarely applied to survival analysis. Next, to derive the consistency of $\hat{\beta}_n$, we define an   integral operator   of   $C(s,t)$ and   derive the consistency based on the compactness of the operator.  When $I(\alpha_0)$ is nonsingular, the accuracy of $\hat{\alpha}_n$  and $\hat{\beta}_n$  can be  measured by the norms $|\cdot|$ and  $\|\cdot\|_{C}$, respectively.

\begin{thm}\label{thm3}
		Suppose that Conditions \ref{assume1}--\ref{assume6} hold. If $I(\alpha_0)$ is nonsingular, then
	\begin{enumerate}[label=(\roman*)]
		\item (Consistency)   $\|\hat{g}_n-g_0\|_{\infty}+\|\dot{\hat{g}}_{n}-\dot{g}_0\|_{\infty}=o_p(1)$,  $|\hat{\alpha}_{n}-\alpha_0|=o_p(1)$, and $\|\hat{\beta}_{n}-\beta_0\|_{C}=o_p(1).$
		\item\label{rate1} (Convergence rate) Let $c=\min\{\omega\nu, \kappa q, (1-\max\{\nu,q\})/2\}.$ We have $$ d(\hat{\xi}_n,\xi_0)=O_p(n^{-c}).$$
		\item\label{rate2} $
			|\hat{\alpha}_n-\alpha_0|+\|\hat{\beta}_n-\beta_0\|_{C}+\|\hat{g}_n-g_0\|_{\mathcal{G}}=O_p(n^{-c}).
$
	\end{enumerate}
\end{thm}
 When $\nu=1/(1+2\omega)$ and $q=1/(1+2\kappa)$,  Theorem \ref{thm3} \ref{rate1}  implies that the  convergence rate of the sieve estimator $\hat{\xi}_n$ could reach the slower rate between $ n^{\omega/(1+2\omega)}$ and $n^{ \kappa/(1+2\kappa)}$. Combining the derivation with \ref{rate2} can easily show that the functional parameter with the weaker smoothness property  could reach  the optimal rate in nonparametric regression, as given in \cite{Stone1982}. Next, we derive the convergence rate of scalar estimator $\hat{\alpha}_n$ and show it could reach $\sqrt{n}$.

\begin{thm}\label{thm4}
	Suppose that Conditions \ref{assume1}--\ref{assume7} hold and  the information bound  $I(\alpha_0)$ is nonsingular. Let $c=\min\{\omega\nu, \kappa q, (1-\max\{\nu,q\})/2\}.$  When $\nu$ and $q$ satisfy $c-q>\frac{1}{4}$, we have
	\begin{equation*}
\sqrt{n}(\hat{\alpha}_{n}-\alpha_{0}) \stackrel{D}{\longrightarrow} N(0,   \Sigma ),
\end{equation*}
where  $ \Sigma =I(\alpha_{0})^{-1}$ and $\stackrel{D}{\longrightarrow}$ denotes convergence in distribution.
\end{thm}

The above result shows that  $\hat{\alpha}_n$ achieves the information  bound. Therefore, it is asymptotically efficient among all the regular estimators. In Theorem \ref{thm4}, the condition on $\nu$ and $q$ is relatively mild and can be conveniently satisfied in most cases. For example, when  $\nu=q$ and $m=\kappa$, the condition is satisfied if  $\frac{1}{2(1+\kappa)}< q< \frac{1}{2\kappa}$ for $\kappa\geq 3$.

\section{Simulation Studies}\label{sec:numerical studies}

We conducted simulation studies to evaluate the finite-sample performance of the proposed method. For generating the functional covariate, we considered a similar setup as \cite{Qu2016} and  defined $Z(\cdot)$  as
\[
Z(s) = \sum^{50}_{k=1} \xi_k U_k \phi_k (s),
\]
where the $U_k$'s were independently sampled from the uniform distribution on $[-1,1]$, $\xi_k= (-1)^{k+1} k^{-1/2}$, $\phi_1 = 1$, and $\phi_{k+1}(s)=\sqrt{2} \cos (k\pi s)$ for $k\ge 1$. The functional coefficient $\beta_0$ was defined as
\[
\beta_0(s) = \sum^{50}_{i=1} (-1)^k k^{-3/2} \phi_k (s).
\]

The scalar covariates $X_1$ followed $Bernoulli(0.5)$ and $X_2$ followed $N(0,0.5)$ truncated at $\pm 2$.
The transformed failure time $T$ was generated from the following functional accelerated failure time model{:}
$$
	T = X_1 + X_2 +\int_{\setI_0}\beta_{0}(s)Z(s)ds + \varepsilon.
$$
We considered three cases for the error term $\varepsilon$: i) $exp(\varepsilon)\sim Exponential(1)$, ii) $\varepsilon\sim 0.5N(0,1)+0.5N(0,3^2)$,
and iii) the standard extreme-value distribution.
We generated censoring time $\tilde{C}$ from $Uniform[0,\tau]$, where $\tau$ was chosen to produce desired censoring rates. Hence the transformed observation time is $Y = \min\{T, \log(\tilde{C})\}$. We considered censoring rates $25\%$ and $40\%$ and sample sizes $n= 400, 600$, and 800. To estimate the functional coefficients $\beta(\cdot)$ and $g(\cdot)$, we adopted the B-spline functions with equally spaced interior knots at the order of $q_n=\lfloor n^{1/4} \rfloor$, resulting in four basis functions for $n=400$ and $600$ and five basis functions for $n=800$, which correspond
to two interior knots for $n=400$ and $600$ and three interior knots for $n=800$.

Let $\{\psi_k(\cdot), k=1, \ldots, q_n\}$ be the spline basis function with support on $\setI_0=[0, 1]$, and
let $\{\eta_k(\cdot), k=1, \ldots, q_n\}$ be the  spline basis function with support on $[a, b]$. The functional parameters $\beta(\cdot)$ and $g(\cdot)$ were approximated by $\sum^{q_n}_{k=1} \beta_k\psi_k(\cdot)$ and $\sum^{q_n}_{k=1} g_k\eta_k(\cdot)$, respectively. The parameter $\tilde\xi=(\alpha_1, \alpha_2,\beta_1,\ldots,\beta_{q_n}, g_1,\ldots,g_{q_n})$ was then estimated based on the following log-likelihood function,
\begin{equation*}
\tilde{l}_n(\tilde \xi)=\frac{1}{n}\sum_{i=1}^{n}\bigg\{\Delta_{i} \sum_{k=1}^{q_n}g_k\eta_k\big(Y_i-\tilde\mu( U_i,\tilde{\theta})\big)-\int e^{\sum_{k=1}^{q_n}g_k\eta_k(t)}I\left\{  Y_i-\tilde\mu(U_i,\tilde{\theta}) \geq t\right\}dt\bigg\},
\end{equation*}
where $\tilde\mu(U,\tilde{\theta})=\alpha^{\top}X+\int_{\setI_0}\sum_{k=1}^{q_n}\beta_k\psi_k(s)Z(s)ds$. Note that the chosen support $[a,b]$ is wide enough such that it covers all residual terms, $Y_i - \tilde{\mu}(U_i,\tilde{\theta})$ for $i=1,\ldots,n$. The standard errors of $\hat\alpha_1$ and $\hat\alpha_2$ were obtained from the first two diagonal entries of $\sqrt{H^{-1}/n}$, where $H$ is the Hessian matrix of $\tilde{l}_n$. For each combination of error distribution, censoring rate and sample size, the simulation was repeated $1000$ times.

\begin{figure}[H]
	\centering
	\includegraphics[width=6in]{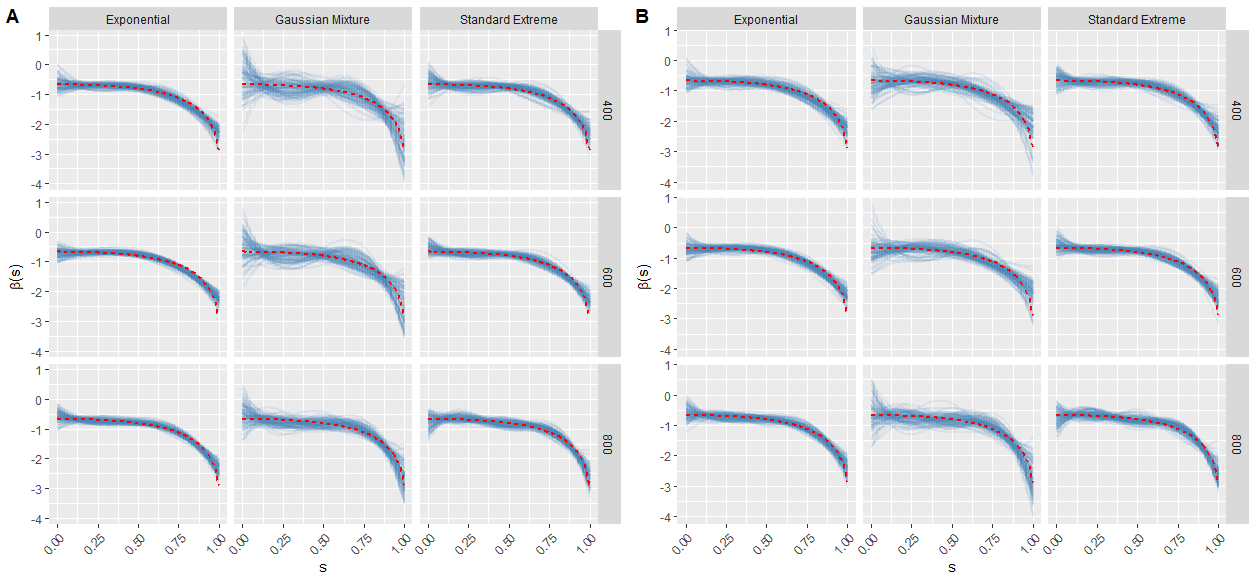}
	\caption{\footnotesize Graphical displays of $\hat{\beta}(\cdot)$.
		The {dashed} lines represent $\beta (\cdot)$ whereas the solid lines represent $\hat{\beta}(\cdot)$. The censoring rate of Panel A and B are 25\% and 40\%, respectively.  \label{beta_example}}	
\end{figure}

\begin{figure}[H]
	\centering
	\includegraphics[width=6in]{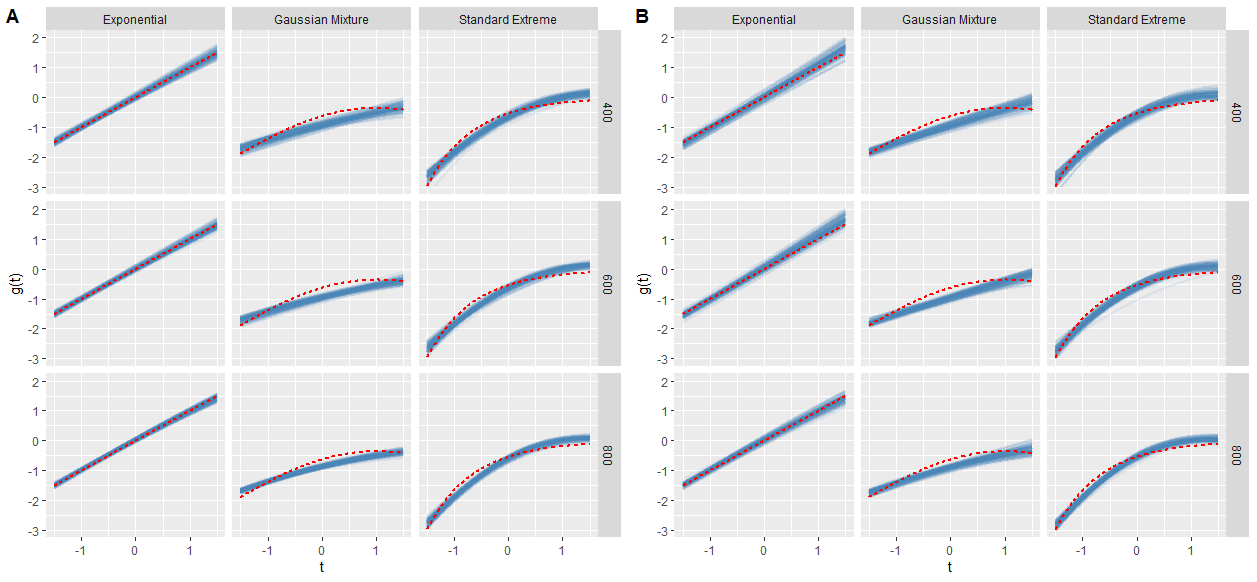}
	\caption{\footnotesize Graphical displays of $\hat{g}(\cdot)$. The {dashed} lines represent $g(\cdot)$ whereas the solid lines represent $\hat{g}(\cdot)$. The censoring rate of Panel A and B are 25\% and 40\%, respectively.\label{g_example}}
\end{figure}

\begin{figure}[H]
	\centering
	\includegraphics[width=6in]{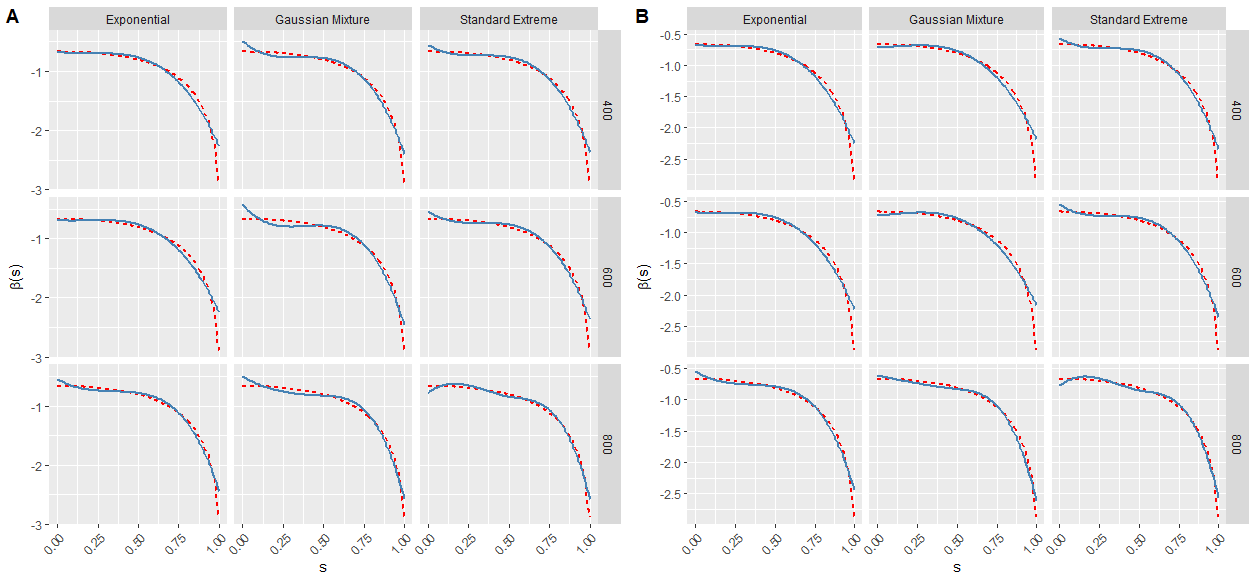}
	\caption{\footnotesize Graphical displays of the pointwise averages $\hat{\beta}(\cdot)$. The {dashed} lines represent $\beta (\cdot)$ whereas the solid lines represent the pointwise averages of $\hat{\beta}(\cdot)$.  The censoring rate of Panel A and B are 25\% and 40\%, respectively.\label{b1}}
\end{figure}

\begin{figure}[H]
	\centering
	\includegraphics[width=6in]{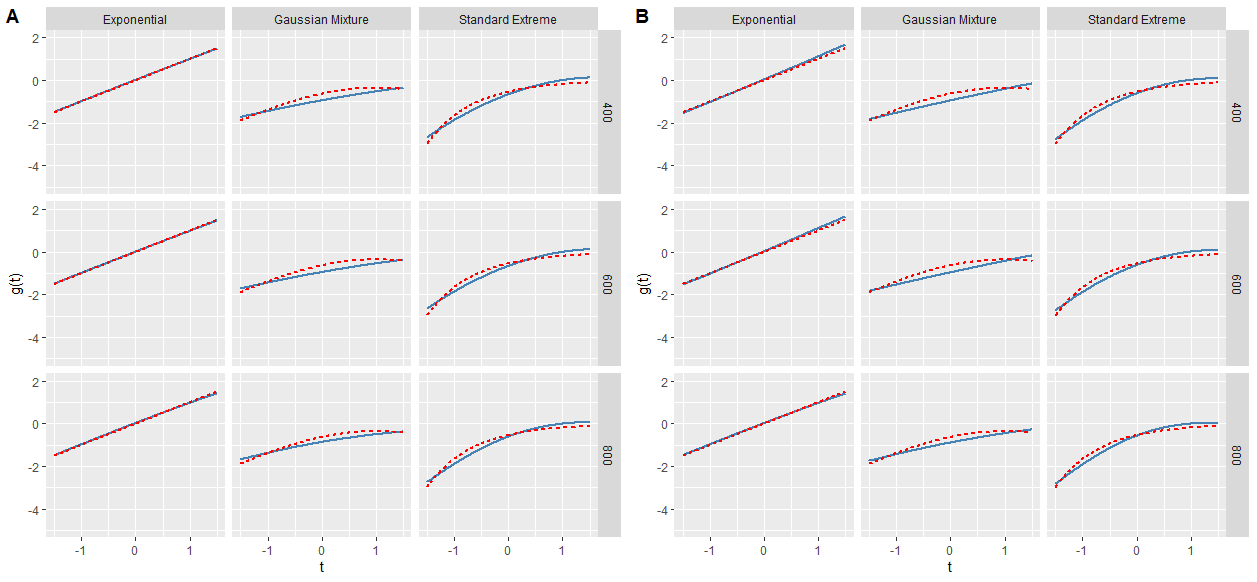}
	\caption{\footnotesize Graphical displays of the pointwise averages $\hat{g}(\cdot)$. The {dashed} lines represent $g(\cdot)$ whereas the solid lines represent the pointwise averages of $\hat{g}(\cdot)$.
		The censoring rate of Panel A and B are 25\% and 40\%, respectively.\label{b2}}
\end{figure}

Figures \ref{beta_example} and \ref{g_example} exhibit 100 instances of estimated $\beta(\cdot)$ and $g(\cdot)$, which mostly clustered around the true values. Figures \ref{b1} and \ref{b2} show the pointwise averages based on 1000 simulations, where the estimates are within close proximity of the true values for both censoring rates under three error distributions.
For the regression coefficient of the scalar covariate, we report the average bias, the sample standard error (SSE), the estimated standard error (ESE), and the coverage probability (CP) in Table \ref{table_1}. Evidently, both SSE and ESE decrease with larger sample sizes and lower censoring rates. Moreover, bias is negligibly small and CP approximates the theoretical level of 95\% across all simulation scenarios. Table \ref{theta_bias} shows the mean squared error (MSE) of parameter estimates $\hat{\beta}(\cdot)$ and $\hat{g}(\cdot)$, which are defined as follows,
\begin{equation*}
	{\rm MSE}(\hat{\beta})={\int_{\setI_0}(\hat{\beta}(s)-\beta_{0}(s))^2ds}
 \qquad \mbox{and} \qquad
{\rm MSE}(\hat{g})={\int^{1.5}_{-1.5}(\hat{g}(t)-g_{0}(t))^2 dt}.
\end{equation*}
For each censoring rate and error distribution, MSE is quite modest for both $\beta(\cdot)$ and $g(\cdot)$, which obviously decreased as sample size increased.
Overall, simulation results validate that both the scalar and functional parameter estimators are consistent and the proposed variance estimation procedure provides reasonable estimates. Furthermore, the empirical coverage probabilities are close to the theoretical level 95\% verifying the normal approximation is appropriate.

\section{Application}\label{sec:application}
As an illustration,
we apply the proposed FAFT model to analyze data from the Improving Care of Acute Lung Injury Patients (ICAP) study (Needham et al., 2006), which is a prospective cohort with a primary goal of assessing longterm outcomes of lower tidal volume ventilation (LTVV) treatments on patients who suffer from acute lung injury/acute respiratory distress syndrome (ALI/ARDS). The study enrolled 520 patients over a two-year period and 413 subjects remained available for analysis after excluding those who stayed in the intensive care unit (ICU) for five days or less, where 153 ($37\%$) died before being discharged from ICU. Among the 413 subjects, the average age was 52 years and 57\% were male.

The sequential organ failure assessment (SOFA) score evaluates the overall health condition of a patient based on dysfunction severity of six organ systems including cardiovascular, central nervous system, coagulation, liver, renal, and respiration. Previous studies have demonstrated that SOFA is a reasonable predictor of outcome prolonged hospital length of stay and mortality of critically ill patients in an ICU (Vincent et al., 1998; Ferreira et al., 2001;  Elias et al., 2020). In the ICAP study, the SOFA score of each patient was recorded upon admission to ICU and daily subsequently during his or her stay. We estimate the functional association between five-day SOFA scores and mortality after controlling for demographical characteristics for patients hospitalized in ICU. The dataset {\it sofa} is available from R package {\it `refund'} (Goldsmith et al., 2021).

\begin{figure}[H]
	\centering
	\includegraphics[width=4in]{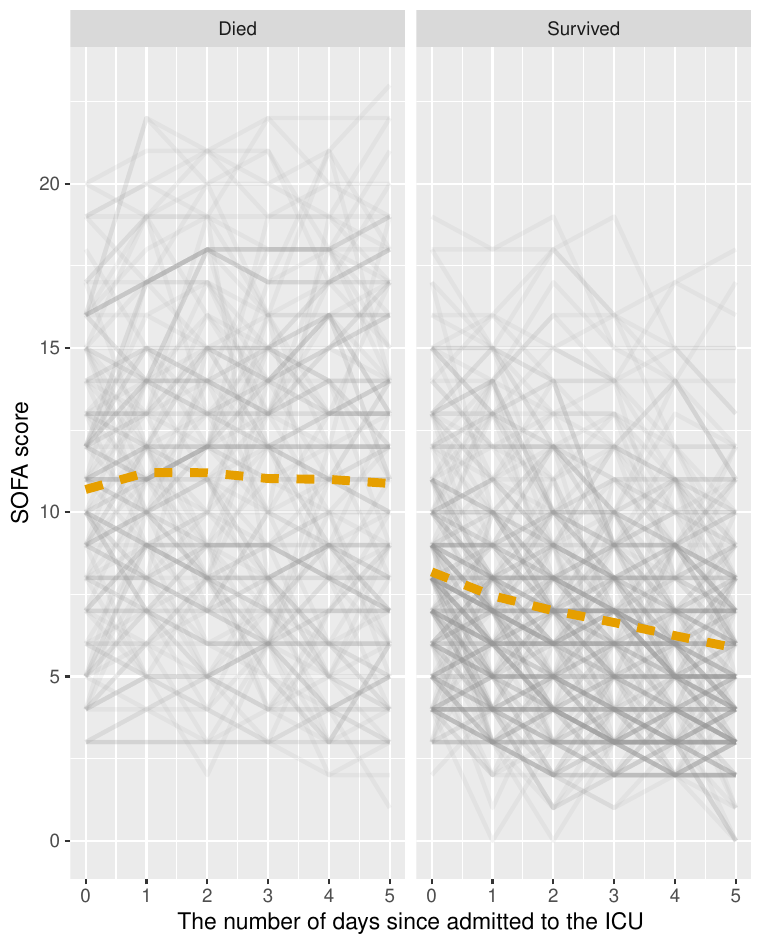}
	\caption{\footnotesize Trajectories of the SOFA score of subjects who died {after the first week} of the ICU hospitalization and those who survived. The orange dotted lines are the pointwise average of the SOFA score.\label{sofax}}
\end{figure}

While the majority analytical methods typically consolidate SOFA scores over a certain time period into discrete values, much information would be lost during such process. The proposed FAFT model offers a convenient alternative that can preserve all the available data on the SOFA score by treating it as a functional covariate. In particular, we treated the SOFA score data from the first 5 days during
the ICU stay as a functional covariate and applied linear interpolation between adjacent measurements. In addition, we considered two scalar covariates, including patient gender and age recorded in years. The event of interest would be death during
the subject ICU stay with the event time calculated as the number of days elapsed after the fifth day in
the ICU till death. For instance, the event time is two days for a patient who died on the seventh day since admission to
the ICU. A patient was considered censored when discharged from the ICU. Figure \ref{sofax} displays trajectories of the 5-day SOFA score for patients who died in the ICU and those discharged from the ICU, where the dotted orange curve represents the pointwise average SOFA scores. Evidently, the trajectory of SOFA scores for patients who died during the ICU stay remains flat and is higher on average compared to those discharged from
the ICU, where an obvious downward trend is observed.

\begin{figure}[H]
	\centering
	\includegraphics[width=4in]{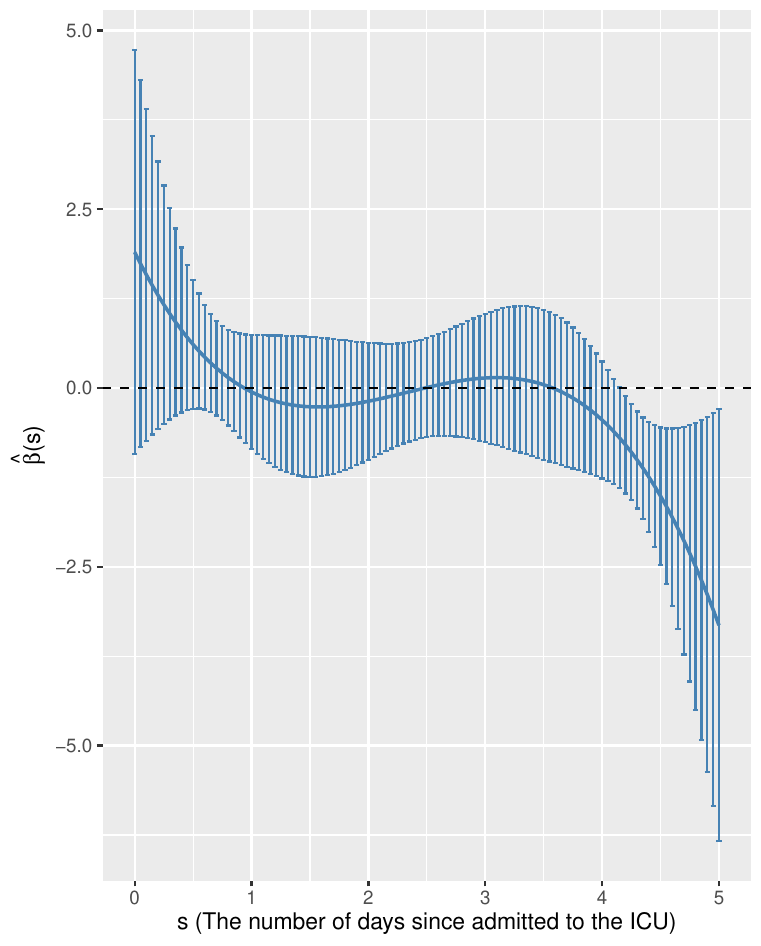}
	\caption{\footnotesize The estimated functional coefficient $\hat{\beta}(\cdot)$ and the pointwise 95\% confidence interval for the SOFA data analysis.\label{sofa}}
\end{figure}

In order to satisfy the condition (A2), we centered subject age and SOFA scores by daily averages as well as assigning values 1 and $-1$ to males and females, respectively. For example, if the average SOFA score in day five is 8, then all SOFA scores in day five would be subtracted by 8. The centered SOFA scores were linearly interpolated using adjacent measurements. We adopted cubic spline functions to estimate the functional coefficient with the basis function at the order of $q_n=\lfloor n^{1/4} \rfloor$ and equally spaced knots. The transformed event time was the natural logarithm of the number of days until death since the fifth day in the ICU. Figure \ref{sofa} shows the estimated functional coefficient $\hat{\beta}(\cdot)$ and the corresponding 95\% pointwise confidence interval, which indicates some functional association between SOFA scores and patient mortality. In particular, a high SOFA score on the fifth day was associated with higher risk of death. Table \ref{sofa_theta} summarizes estimation results for gender and age, where older patients and female patients were associated with higher risk of death.

\section{Concluding Remarks}\label{sec:conclusion}

We studied a functional accelerated failure time model to reveal the effect of a functional covariate on survival time as well as provided sufficient conditions to guarantee the model identifiability. Under this model, the likelihood function exhibits a complex structure where parametric and nonparametric coefficients are bundled inside an unknown baseline hazard function. To overcome the challenges arising from the bundled parameters, we developed a sieve-based maximum likelihood estimation procedure. Subsequently, we derived the convergence rate of the proposed estimators which achieved the optimal rate under the framework of semiparametric inference as well as establishing the semiparametric efficiency of the finite-dimensional estimator. Simulations under various settings and a real application both yielded satisfactory results, which indicate good performance of the proposed efficient estimation approach. 

The proposed approach can be extended to making inference for the partial linear accelerated failure time model and other semiparametric survival models such as functional Cox models and functional additive hazards models. Recently, Zhong, Mueller and Wang (2021, 2021+) developed the deep learning approaches for statistical inference based on the partially linear Cox model and a general class of hazard models. Another interesting research topic is to investigate functional survival models such as functional AFT and Cox models using deep natural networks.

\pagebreak

\appendix
\section{Proofs of Theorems}\label{app-A}
\subsection{Lemmas}
First, we introduce notation. Let the symbol  $\lesssim$   denote that the left-hand  side  is bounded above   by a constant multiplied with the right-hand side, and let the symbol $\succsim $ denote   that the left-hand side  is bounded below  by a constant multiplied with the right-hand side. Let $|\cdot|$ denote the Euclidean norm, and let $\|\cdot\|$ denote the $L_2$-norm with respect to a probability  measure that should be clear in the context. Also,  let  $\|\cdot\|_{\infty}$ denote the supremum norm. Let $P$ and $\mathbb{P}_{n}$  denote the probability measure and the empirical measure, respectively.  Let $\mathbb{G}_{n}=\sqrt{n}(\mathbb{P}_{n}-P)$ denote the empirical process.
For a twice differentiable function $f$, we denote its first derivative and second derivative  by   $\dot{f}$  and   $\ddot{f}$, respectively. Let   $\mathcal{W}=(Y,\Delta,U)$.    Define $r_{\theta}=Y-\mu(U,\theta)$ and  $t_{\theta}=t-\mu(U,\theta-\theta_{0})$. Then, we have $r_{\theta_{0}}=Y-\mu(U,\theta_{0})=\min\{\varepsilon,C-\mu(U,\theta_{0})\}$ and $t_{\theta_{0}}=t$.
\vspace{0.2in}

 Under Conditions \ref{assume1}--\ref{assume4},   the log-likelihood function for a single sample $\mathcal{W}$  is
	\begin{equation*}
  m(\xi;\mathcal{W})=\Delta g\left(r_{\theta}\right)-\int_{a}^{b}I\left\{ r_{\theta_{0}}\geq t\right\}\exp\left\{g\left(  t_{\theta}\right)\right\}dt,
	\end{equation*}
up to an additive term that is not dependent on $\xi$. Consider a smooth and parametric submodel $\big\{\xi_{(\varepsilon)}=(\alpha+\varepsilon h_1, \beta+\varepsilon h_2, g+\varepsilon h_3): \varepsilon$ in  a neighborhood of  $0\in\mathbb{R}\} $, which  satisfies  $\xi_{(\varepsilon)}\mid_{\varepsilon=0}=\xi$.
We denote the derivative of the log-likelihood $m(\xi;\mathcal{W})$ with respect to the direction $\bh=(h_1,h_2,h_3)$  by
	\begin{equation*}
	\begin{split}
		\dot{m}(\xi;\mathcal{W})[\bh]&=\frac{\partial m(\xi_{(\varepsilon)};\mathcal{W})}{\partial \varepsilon}\bigg|_{\varepsilon=0}
		\\& =\bigg\{\int_{a}^{b}I(r_{\theta_{0}}\geq t )  \exp\{g(t_{\theta})\}\dot{g}(t_{\theta})
		dt
		-\Delta \dot{g}(r_{\theta})\bigg\}	\mu(U,\delta)
		\\&\quad+\Delta h_3(r_{\theta})    -\int_{a}^{b}I(r_{\theta_{0}}\geq t) \exp \{g(t_{\theta})\} h_3(t_{\theta})dt,
	\end{split}
\end{equation*}
where $\delta=(h_1, h_2)$ and $\mu(U,\delta)= h_1^{\top}X+\int_{0}^{1}h_2(s)Z(s)ds$.
Define  $\mathcal{H}_{\beta}=\big\{h_2\in\mathcal{F}^{\omega}: P\big[\Delta\big(\int_{0}^{1}h_2(s)Z(s)ds\big)^2\big]<\infty\big \}$ and $\mathcal{H}_{g}=\big\{h_3\in \mathcal{G}^{\kappa-1}: P\big[\Delta h_3(r_{\theta_0})^2\big]<\infty\big\}$. Let    $\mathcal{H}=\big\{(h_1,h_2, h_3): h_1\in\setR^{p},  h_2\in\mathcal{H}_{\beta}, h_3\in\mathcal{H}_{g}\big\}$.
 For  $\tilde{\bh}=(\tilde{h}_1,\tilde{h}_2,\tilde{h}_3)\in \mathcal{H},$ we denote
	\begin{equation*}
		\begin{split}
		&\;	\ddot{m}(\xi;\mathcal{W})[\bh,\tilde{\bh}]
		\\&=\frac{\partial \dot{m}(\alpha+\varepsilon \tilde{h}_1, \beta+\varepsilon \tilde{h}_2, g+\varepsilon \tilde{h}_3;\mathcal{W})[\textbf{h}]}{\partial \varepsilon}\bigg|_{\varepsilon=0}
			\\&=  \bigg\{-\int_{a}^{b}I(r_{\theta_{0}}\geq t )\exp \{g(t_{\theta})\}\{ \ddot{g} +(\dot{g})^2\}(t_{\theta}) dt
			+ \Delta \ddot{g}(r_{\theta})\bigg\}\mu(U,\delta)\mu(U,\tilde{\delta})
			\\&\quad +\bigg\{\int_{a}^{b}I(r_{\theta_{0}}\geq t )\exp \{g(t_{\theta})\}\{\dot{\tilde{h}}_3+\dot{g}\tilde{h}_3\}(t_{\theta})dt
			- \Delta \dot{\tilde{h}}_3(r_{\theta})\bigg\}\mu(U,\delta)
			\\&\quad +\bigg\{\int_{a}^{b}I(r_{\theta_{0}}\geq t )\exp \{g(t_{\theta})\}\{\dot{h}_3+\dot{g}h_3\}(t_{\theta})dt
			-\Delta \dot{h}_3(r_{\theta})\bigg\}\mu(U,\tilde{\delta})
			\\&\quad- \int_{a}^{b}I(r_{\theta_{0}}\geq t) \exp\left\{g(t_{\theta}) \right\} h_3(t_{\theta})\tilde{h}_3(t_{\theta})dt,
		\end{split}
	\end{equation*}
	where $\tilde{\delta}=(\tilde{h}_1, \tilde{h}_2)$ and  $\mu(U,\tilde{\delta})= \tilde{h}_1^{\top}X+\int_{0}^{1}\tilde{h}_2(s)Z(s)ds.$
\vspace{0.5cm}

The following Lemma is  a direct result according to Corollary 6.21 of \cite{Schumaker1981}.
\begin{lemma}\label{lem:appro}
Under Conditions \ref{assume4} and \ref{assume5}, for any $\beta_0\in\mathcal{F}^{\omega}$ and $g_0\in\mathcal{G}^{\kappa}$, there exist  functions   $\beta_{0n}\in\mathcal{F}_{n}^{\omega}$ and $g_{0n}\in\mathcal{G}^{\kappa}_{n}$  such that
\begin{equation*}
	\|\beta_{0n}-\beta_{0}\|_{\infty}=O(n^{-\omega \nu})  \text{ and } \|g_{0n}-g_{0}\|_{\infty}=O(n^{-\kappa q}).
\end{equation*}
 Define $\xi_{0n}=(\alpha_{0}, \beta_{0n}, g_{0n})$. Then, $\xi_{0n}$ belongs to  $ \Xi_n$ and   satisfies $\|\xi_{0n}-\xi_{0}\|_{\infty}=O(n^{-\min\{\omega \nu, \kappa q\}})$.
\end{lemma}

\begin{lemma}\label{lem3}
 Denote a class of functions $ \mathcal{F}_{n}=\left\{m(\xi;\mathcal{W})-m(\xi_{0};\mathcal{W}):\xi\in\Xi_{n}\right\}.$  Suppose that   Conditions  \ref{assume1}--\ref{assume5} hold,   the $\varepsilon$-bracketing number associated with $\|\cdot\|_{\infty}$ for $\mathcal{F}_{n}$ satisfies
\begin{equation*}
	N_{ [\;] }(\varepsilon,\mathcal{F}_{n},\|\cdot\|_{\infty})\lesssim (1/\varepsilon)^{c_1m_{n}^{\omega}+c_2s_{n}^{\kappa}+p},
\end{equation*}
for some   constants $c_1,c_2>0$.
\end{lemma}
\begin{lemma}\label{lem4}
Define a class of functions
	\begin{equation*}
		\begin{split}
			\mathcal{F}_{n}(\bh,\eta)=\{\dot{m}(\xi;\mathcal{W})[\bh]-\dot{m}(\xi_{0};\mathcal{W})[\bh]:\;& \xi\in\Xi_{n} \text{ and } d(\xi,\xi_{0})\leq \eta \,;\\&\dot{g}\in\mathcal{G}_{n}^{\kappa-1}  \text{ and } \|\dot{g}-\dot{g}_{0}\|_{\mathcal{G}}\leq \eta\}.
		\end{split}
	\end{equation*}
	Suppose that  Conditions  \ref{assume1}--\ref{assume5} hold, then
	$$N_{[\;]}(\varepsilon, \mathcal{F}_{n}(\bh,\eta),\|\cdot\|_{\infty})\lesssim (\eta/\varepsilon)^{c_3m_{n}^{\omega}+c_4s_{n}^{\kappa}+p}$$ for some   constants $c_3,c_4>0.$
\end{lemma}
\begin{lemma}\label{lem5}
Define a sequence of space $\mathcal{H}_{n}(\bh,\eta)=\big\{(h_{1},\tilde{h}_2, \tilde{h}_3):   \tilde{h}_2\in\mathcal{F}_{n}^{\omega},  \tilde{h}_3\in\mathcal{G}_{n}^{\kappa} \text{ such that } \|\tilde{h}_2-h_2\|_{\infty}\leq \eta, \|\tilde{h}_3-h_3\|_{\infty}\leq \eta \big\}$.  Denote a class of functions
	\begin{equation*}
		\begin{split}
			\mathcal{L}_{n}(\bh,\eta)=\Big\{\dot{m}(\xi;\mathcal{W})[\bh-\tilde{\bh} ]:\; &\xi\in\Xi_{n}\text{ and } d(\xi,\xi_{0})\leq \eta;\; \dot{g}\in\mathcal{G}_{n}^{\kappa-1}\\& \text{ and }   \|\dot{g}-\dot{g}_{0}\|_{\mathcal{G}}\leq \eta;\; \tilde{\bh}\in\mathcal{H}_{n}(\bh,\eta) 	 \Big\}.
		\end{split}
	\end{equation*}
Suppose that   Conditions  \ref{assume1}--\ref{assume5} hold. Then,
 \begin{equation*}
	N_{ [\;] }(\varepsilon,\mathcal{L}_{n}(\bh,\eta),\|\cdot\|_{\infty})\lesssim (\eta/\varepsilon)^{c_5m_{n}^{\omega}+c_6s_{n}^{\kappa}+p},
\end{equation*}
for some  constants $c_5, c_6>0.$
\end{lemma}

\subsection{Proof of Proposition  1}
\begin{proof}
	The score vector for $\alpha_0$, the score operator for  functional parameter $\beta_0$, and the score operator  for  log-hazard function $g_0$ are given as
	\begin{equation*}
	\begin{split}
	&\dot{l}_{\alpha_0}=-X\int \dot{g}_0(t)dM(t),
	\\&\dot{l}_{\beta_0}h_{2}= -\int_0^1h_2(s)Z(s)ds \int \dot{g}_0(t)dM(t),
	\\& \dot{l}_{g_0}h_{3}=\int h_3(t)dM(t),
	\end{split}
	\end{equation*}
	where $M(t)=\Delta I(r_{\theta_{0}}\leq t)-\int^{t}_{-\infty}I(r_{\theta_{0}}\geq u)\lambda_{0}(u)du$ is a counting process martingale.
	The efficient score function for $\alpha_0$ is defined by $\dot{l}^{\star}_{\alpha_0}=\dot{l}_{\alpha_0}-\dot{l}_{\beta_0} b^{\star}-\dot{l}_{g_0}\phi^{\star},$ where $b^{\star} =(h_{21}^{\star},\dots,\hs_{2p})^{\top}$ and $\phi^{\star}=(\hs_{31},\dots,\hs_{3p})^{\top}$ are the directions such that $\dot{l}^{\star}_{\alpha_0}$  is orthogonal to the nuisance tangent space,  given as $\textbf{L}= \overline{Span}\{\dot{l}_{\beta_0}h_2+\dot{l}_{g_0}h_3: h_2\in\mathcal{H}_{\beta} \text{ and } h_3 \in\mathcal{H}_{g}\}$. Based on \cite{Huang1999}, it suffices to solve the following minimization problem.
	\begin{equation}\label{eq:pro}
	\min P\Big[\Delta\Big|\dot{l}_{\alpha_0}-\dot{l}_{\beta_0} b -\dot{l}_{g_0} \phi \Big|^2\;\Big].
	\end{equation}
	According to Conditions \ref{assume2}--\ref{assume4} and \ref{assume6},  the solution to \eqref{eq:pro}  is well defined. One choice for  $\phi^{\star}$ is given by
	\begin{equation*}
	\phi^{\star}(t)=-\dot{g}_{0}(t)P\big(X- \int_{0}^{1} b^{\star}(s)Z(s)ds \big| \varepsilon= t, \Delta=1\big).
	\end{equation*}
	By  Conditions \ref{assume7}, it can be shown that  each element of  $\phi^{\star}$  is  $(\kappa-1)$-th differentiable and   has bounded $(\kappa-1)$-th derivative.  The information bound for estimation of $\alpha_0$ is
	\begin{equation*}
	I(\alpha_0)=P \big[\dot{l}^{\star\otimes2}_{\alpha_0}\big]=P\Big[\Delta \big(-\dot{g}_0(r_{\theta_0}) X + \dot{g}_0(r_{\theta_0}) \int_0^1 b^{\star}(s)Z(s)ds- \phi^{\star}( r_{\theta_0})\big)^{\otimes2}\Big],
	\end{equation*}
	where $x^{\otimes 2}=xx^{\top}$ for any vector $x\in\mathbb{R}^{p}.$
\end{proof}

 \subsection{Proof of Proposition  2}
 \begin{proof}
 	Define
 	$
 	M_{n}(\xi)=\mathbb{P}_{n}m(\xi;\mathcal{W})$ and $ M(\xi)=Pm(\xi;\mathcal{W}).$
 	  Direct calculation yields
 	\begin{equation}\label{M function}
 	\begin{split}
 	M(\xi_{0})-M(\xi)
 	=P\Big(\Delta\phi\big\{g\left(r_{\theta}\right)-g_{0} (r_{\theta_{0}} ) \big\}\Big),
 	\end{split}
 	\end{equation}
 	where $\phi(x)=\exp(x)-1-x.$ The function $\phi$ satisfies $\phi(x)\geq 0$ for any $x$, and  $\phi(x)=0$  only when $x=0$. Hence, for any $\xi\in\Xi$, we have  $M(\xi_{0})\geq M(\xi)$, and $M(\xi_{0})= M(\xi)$ holds only when $\Delta\left\{g\left(r_{\theta}\right)-g_{0} (r_{\theta_{0}} ) \right\}^2=0,$ a.s..   Therefore, $\|g\left(r_{\theta}\right)-g_0(r_{\theta_{0}})\|_{\mathcal{A}}=0$ is shown.
 	
 	Next, we study separate parameter.  We generate $V$ as an independent copy of $U$. Suppose that $P\big\{\mu(U,\theta-\theta_0)^2\big\}>0$ holds. We   have $P\big(\big\{\mu(U,\theta-\theta_0)-\mu(V,\theta-\theta_0)\big\}^2\big)>0$.
 	By decomposing the image space, we can derive    a sequence of nested closed interval $\big\{ [r_n,l_n]\big\}_{n\geq 1}$ such that $[r_n,l_n]\subseteq [r_{n-1},l_{n-1}]$, $|r_n-l_n|\leq 2^{-n}c_1$ for some constant  $c_1$, and $E(I_{A_n})>0$, where $A_n = \big\{ r_n\leq \mu(U,\theta-\theta_{0})-\mu(V,\theta-\theta_{0})\leq l_n\big\}$.    The design implies that  $\{r_n\}_{n\geq 1}$ is a Cauchy sequence, hence  there is $r^{\star}$ such that $r_n \rightarrow r^{\star}$. Based on  Condition \ref{assume3},  we can derive
 	\begin{equation*}
 	\begin{split}
 	\|g\left(r_{\theta}\right)-g_0(r_{\theta_{0}})\|_{\mathcal{A}}^{2}
 	&=P\left[\int_{a}^{b}P\big( r_{\theta_{0}}\geq t  |U\big)\exp\{g_0(t)\}\big(g\left(t_{\theta}\right)-g_0(t)\big)^{2}dt\right]
 	\\&\geq   \delta  \int_{a}^{b}\exp\{g_0(t)\}P\big( g\left(t_{\theta}\right)-g_0(t)\big)^2dt.
 	\end{split}
 	\end{equation*}
 	It implies that when $\|g\left(r_{\theta}\right)-g_0(r_{\theta_{0}})\|_{\mathcal{A}}=0$  we have $g(t_{\theta})=g_{0}(t)$, a.s.. Hence,   $g(t-\mu(U,\theta-\theta_0))=g_{0}(t)$ and $g(t-r^{\star}-\mu(V,\theta-\theta_0))=g_{0}(t-r^{\star})$ hold, a.s.,  which follows that
 	\begin{equation*}
 	\begin{split}
 	\big|g_{0}(t)-g_{0}(t-r^{\star})\big|&=\big|g(t-\mu(U,\theta-\theta_0))-g(t-r^{\star}-\mu(V,\theta-\theta_0))\big|
 	\\&\leq \sup|\dot{g}|\cdot \big|\mu(U,\theta-\theta_0)-\mu(V,\theta-\theta_0)-r^{\star}\big|
 	\\&\leq \sup|\dot{g}|\cdot \left(|\mu(U,\theta-\theta_0)-\mu(V,\theta-\theta_0)-r_n|+|r_n-r^{\star}|\right).
 	\end{split}
 	\end{equation*}
 	By multiplying  $I_{A_n}$ on the both side and taking expectation, we get
 	\begin{equation*}
 	\begin{split}
 	E \big(\big|g_{0}(t)-g_{0}(t-r^{\star})\big|\cdot I_{A_n}\big)
 	&=\big|g_{0}(t)-g_{0}(t-r^{\star})\big|\cdot E ( I_{A_n})
 	\\& \leq \sup|\dot{g}|\cdot \left(|l_n-r_n| +|r_n-r^{\star}|\right)\cdot E (I_{A_n}).
 	\end{split}
 	\end{equation*}
 	Since  $E ( I_{A_n})>0$, it leads to
 	$$\big|g_{0}(t)-g_{0}(t-r^{\star})\big|\leq \sup|\dot{g}|\cdot \left(|l_n-r_n| +|r_n-r^{\star}|\right)\rightarrow 0, \text{  a.s.}.$$
 	This contradicts with the condition that $g_{0}$ is a non-periodic function. Therefore, the assumption is incorrect and $P\{\mu(U,\theta-\theta_0)^2\}=0$ is verified.   Plugging this into $g(t_{\theta})=g_0(t)$, a.s.,  the identifiability of $g_0$  is  also shown. Therefore, we demonstrate that   $d(\xi, \xi_0)=0$ is guaranteed by $\|g\left(r_{\theta}\right)-g_0(r_{\theta_{0}})\|_{\mathcal{A}}=0$.   Moreover,  from  Proposition 1, it shows  that for any $\theta$, we have
 	\begingroup
 	\allowdisplaybreaks
 	\begin{align}\label{dis}
 	&P\Big[\Delta \Big\{\dot{g}_{0}(r_{\theta_{0}})\mu(U,\theta-\theta_0)\Big\}^2\Big]
 	\nonumber
 	\\&=P\Big[\Delta \Big\{\dot{g}_{0}(r_{\theta_{0}}) X^{\top}(\alpha-\alpha_0)+\dot{g}_{0}(r_{\theta_{0}})\int_{0}^{1}Z(s)\big(\beta(s)-\beta_{0}(s)\big) ds \Big\}^2\Big]
 	\nonumber
 	\\&=P\Big[\Delta \Big\{\big(\dot{g}_{0}(r_{\theta_{0}})X-\eta(r_{\theta_0};b^{\star},\phi^{\star})\big)^{\top}(\alpha-\alpha_0)+\eta(r_{\theta_0};b^{\star},\phi^{\star})^{\top}(\alpha-\alpha_0) \nonumber
 	\\&\qquad\quad +\dot{g}_{0}(r_{\theta_{0}})\int_{0}^{1}Z(s)\big(\beta(s)-\beta_{0}(s)\big) ds \Big\}^2\Big]
 	\nonumber
 	\\&=P\Big[\Delta \Big\{\big(\dot{g}_{0}(r_{\theta_{0}})X-\eta(r_{\theta_0};b^{\star},\phi^{\star})\big)^{\top}(\alpha-\alpha_0)\Big\}^2\Big]
 	\nonumber
 	\\&\qquad+P\Big[\Delta \Big\{ \eta(r_{\theta_0};b^{\star},\phi^{\star})^{\top}(\alpha-\alpha_0)+\dot{g}_{0}(r_{\theta_{0}})\int_{0}^{1}Z(s)\big(\beta(s)-\beta_{0}(s)\big) ds \Big\}^2\Big],
 	\end{align}	
 	\endgroup
 	where $\eta(t;b^{\star},\phi^{\star})= \dot{g}_{0}(t)\int_{0}^{1} b^{\star}(s)Z(s)ds- \phi^{\star}(t).$ Therefore, when $I(\alpha_{0})$ is nonsingular, $P\{\mu(U,\theta-\theta_0)^2\}=0$  and \eqref{dis} lead to $|\alpha-\alpha_0|=0$, which in turn implies $\|\beta-\beta_{0}\|_{C}=0$.
 \end{proof}

 \subsection{Proof of Theorem 1}
 \begin{proof}

\medskip

  By calculation,   $\phi(x)\geq   \frac{1}{2}\exp\{-c_n\}x^2$ holds when $x\in[-c_n,c_n]$. Then,    there is a  constant $c$ such that  for any $\xi\in\Xi_{n}$,
\begin{equation*}
	\begin{split}
		M(\xi_{0})-M(\xi)  \geq  c\exp\{-c_n\}P\big(\Delta\left\{ g(r_{\theta})-g_{0} (r_{\theta_{0}} )  \right\}^2\big).
	\end{split}
\end{equation*}
Therefore,
\begin{equation}\label{C1check}
	\inf_{\{\xi\in\Xi_{n}:\;\|g\left(r_{\theta}\right)-g_0(r_{\theta_{0}})\|_{\mathcal{A}}\geq \varepsilon\}}	M(\xi_{0})-	 M(\xi)\geq 2A\exp\{-c_n\}\varepsilon^2,
\end{equation}
for  a constant $A>0$.

\smallskip

By  Lemma  \ref{lem:appro}, there exists  $\xi_{0n}\in\Xi_{n}$  such that $\|\xi_{0}-\xi_{0n}\|_{\infty}=O(n^{-\min\{\omega\nu,\kappa q\}}).$ This together with
$\| m(\xi_{0};\mathcal{W})-m(\xi_{0n};\mathcal{W})\|_{\infty}\lesssim\|\xi_{0}-\xi_{0n}\|_{\infty}$  leads to $\| m(\xi_{0};\mathcal{W})-m(\xi_{0n};\mathcal{W})\|_{\infty}=O(n^{-\min\{\omega\nu,\kappa q\}}).$
Considering a class of functions $\mathcal{F}_{n}=\left\{  m(\xi; \mathcal{W})-m(\xi_{0}; \mathcal{W}):\xi\in\Xi_{n}\right\}$, Lemma \ref{lem3} shows that   the $\varepsilon$-bracketing number associated with the supremum norm for $\mathcal{F}_{n}$ satisfies $
N_{ [\;] }(\varepsilon,\mathcal{F}_{n},\|\cdot\|_{\infty})\lesssim (1/\varepsilon)^{c_1m_{n}^{\omega}+c_2s_{n}^{\kappa}+p},
$ for some   constants $c_1,c_2>0$.  It follows by Theorem 2.5.6 of \cite{Van1996} that
\begin{equation*}
	\underset{\xi\in\Xi_{n}}{\sup} \Big|(\mathbb{P}_{n}-P)\left\{m(\xi; \mathcal{W})-m(\xi_{0}; \mathcal{W})\right\}\Big|=O_p(n^{-1/2}).
\end{equation*}
For each $n\geq 1$,   let $\Xi_{n}(\varepsilon)= \{\xi\in\Xi_{n}:\;\|g\left(r_{\theta}\right)-g_0(r_{\theta_{0}})\|_{\mathcal{A}}\geq \varepsilon\}$.  Since the sieve space $\Xi_{n}$ is  compact  and $M(\xi)$ is continuous on $\Xi_{n}$,  $\sup_{\xi\in\Xi_{n}(\varepsilon)}M(\xi)-M(\xi_{0})$ is bounded. The calculation shows
\begin{equation*}
	\begin{split}
			&\sup_{\xi\in\Xi_{n}(\varepsilon)}M_n(\xi;\mathcal{W})-M_n(\xi_{0n};\mathcal{W})
			\\&\leq
			\sup_{\xi\in\Xi_{n}(\varepsilon)}(\mathbb{P}_n-P)\left\{m(\xi; \mathcal{W})-m(\xi_{0}; \mathcal{W})\right\}-  \left\{ M_n(\xi_{0n}; \mathcal{W})-M_n(\xi_{0}; \mathcal{W})\right\}
		\\&\qquad +\sup_{\xi\in\Xi_{n}(\varepsilon)}M(\xi)-M(\xi_{0})
		\\&\leq O_p(n^{-\min\{\omega\nu,\kappa q,1/2\} })-2A\exp\{-c_n\}\varepsilon^2.
	\end{split}
\end{equation*}
By a similar argument  as the proof of Theorem 3.1 in \cite{Chen2007},  we show that
\begin{equation*}
	  \begin{split}
		P\left( \|\hat{g}_n\left(r_{\hat{\theta}_n}\right)-g_0(r_{\theta_{0}})\|_{\mathcal{A}}\geq \varepsilon\right)&\leq P\Big(\sup_{\xi\in\Xi_{n}(\varepsilon)}M_n(\xi;\mathcal{W})-M_n(\xi_{0n};\mathcal{W})\geq 0\Big)
		\\&\leq P\Big(O_p(n^{-\min\{\omega\nu,\kappa q,1/2\} })\geq 2A\exp\{-c_n\}\varepsilon^2\Big)
		\\&=o(1),
	\end{split}
\end{equation*}
where the last equality  holds since $c_n$ grows with $n$ slowly enough. Therefore, we demonstrate $\|\hat{g}_n\left(r_{\hat{\theta}_n}\right)-g_0(r_{\theta_{0}})\|_{\mathcal{A}}=o_p(1).$
\medskip

To establish the convergence rate,  we  apply the result  of \cite{Shen1994}. The proof proceeds by verifying their conditions  C1--C3.    The  condition C1  of \cite{Shen1994}  is demonstrated by  \eqref{C1check},  with the constant $\alpha=1$ and $A_1=A\exp\{-c_n\}$ in their notation.
 \smallskip

 Next, we consider     the  condition C2 of  \cite{Shen1994}.  Based on Cauchy-Schwarz inequality, it follows that
 \begin{equation*}
 	\begin{split}
 &	\big\{m(\xi;\mathcal{W})-m(\xi_{0};\mathcal{W})\big\}^{2}
 \\&=\left\{\Delta g(r_{\theta})-\Delta g_{0}(r_{\theta_{0}})+\int_{a}^{b}I(r_{\theta_{0}}\geq t)\big(\exp\{g_{0}(t)\}-\exp\{g(t_{\theta})\}\big)dt\right\}^2
   \\&\lesssim \Delta \big\{g(r_{\theta})-  g_{0}(r_{\theta_{0}})   \big\}^2 +\int_{a}^{b}I(r_{\theta_{0}}\geq t)\big(\exp\{g_{0}(t)\}-\exp\{g(t_{\theta})\}\big)^2dt
\\&= I_1+I_2.
 	\end{split}
 \end{equation*}
 For $I_1,$  we obtain
 \begin{equation*}
 	\begin{split}
 		P(I_1)=P\big( \Delta \{g(r_{\theta})-  g_{0}(r_{\theta_{0}})   \}^2\big)
              =\|g\left(r_{\theta}\right)-g_0(r_{\theta_{0}})\|_{\mathcal{A}}^2.
 	\end{split}
 \end{equation*}
 For $I_2$, we first derive that the inequality $|\exp(x)-1|\leq |\exp(|x|)-1|\leq |x|\exp(|x|)$ holds for any $x$. Based on Conditions \ref{assume4} and \ref{assume5}, it follows that
\begin{equation*}
	\begin{split}
	P(I_2)&=P\left[\int_{a}^{b}I(r_{\theta_{0}}\geq t)\big[\exp\{g_{0}(t)\}-\exp\{g(t_{\theta})\}\big]^2dt \right]
	\\&= P\left[\int_{a}^{b}I(r_{\theta_{0}}\geq t)\exp\{2g_{0}(t)\}\big[1-\exp\{g(t_{\theta})-g_{0}(t)\}\big]^2dt \right]
	\\&\leq \|\exp\{2|g|+3|g_{0}|\}\|_{\infty}P\left[\int_{a}^{b}I(r_{\theta_{0}}\geq t)\exp\{g_{0}(t)\}\left\{g(t_{\theta})-g_{0}(t)\right\}^2dt \right]
	\\&\lesssim \exp\{2c_n\} \|g(r_{\theta})-g_0(r_{\theta_{0}})\|_{\mathcal{A}}^2.
	\end{split}
\end{equation*}
 Therefore, for any $\xi\in\Xi_{n},$ we have
\begin{equation*}
	P\big\{m(\xi;\mathcal{W})-m(\xi_{0};\mathcal{W})\big\}^{2}\lesssim \exp\{2c_n\}\|g(r_{\theta})-g_0(r_{\theta_{0}})\|_\mathcal{A}^2,
\end{equation*}
which  implies
\begin{equation*}
	\begin{split}
	\underset{\left\{\xi\in\Xi_{n}:\;\|g(r_{\theta})-g_0(r_{\theta_{0}})\|_\mathcal{A}\leq \varepsilon\right\}}{ \sup}Var \big\{m(\xi;\mathcal{W})-m(\xi_{0};\mathcal{W})\big\}
	\lesssim \exp\{2c_n\}\varepsilon^2.
	\end{split}
\end{equation*}
 This indicates that  the  condition C2  of \cite{Shen1994} holds  with the constant $\beta=1$  and $A_2=C \exp\{2c_n\}$   in their notation, where  $C$ is a constant.

 \smallskip

Finally, we investigate  the  condition C3 of \cite{Shen1994}.  Following  a similar argument   as  that of  Lemma \ref{lem3}, we can show that the $\varepsilon$-bracketing number associated with the supremum norm  for the  class of  functions
$$\widetilde{\mathcal{F}}_{n}=\left\{m(\xi;\mathcal{W})-m(\xi_{0n};\mathcal{W}): \xi\in\Xi_{n}\right\}$$
satisfies $N_{ [\;] }(\varepsilon,\widetilde{\mathcal{F}}_{n},\|\cdot\|_{\infty})\lesssim (1/\varepsilon)^{c_1m_{n}^{\omega}+c_2s_{n}^{\kappa}+p}$ for some constants $c_1,c_2>0$. This implies
 \begin{equation*}
 	\begin{split}
	H(\varepsilon,\widetilde{\mathcal{F}}_{n},\|\cdot\|_{\infty})=\log N(\varepsilon,\widetilde{\mathcal{F}}_{n},\|\cdot\|_{\infty})\lesssim n^{\max\{\nu,q\}}\log(1/\varepsilon),
	\end{split}
 \end{equation*}
which demonstrates that the condition C3 of \cite{Shen1994} holds with the constant $r_{0}=\max\{\nu,q\}/2$ and $r=0^{+}$ in their notation.

Therefore, it follows that the constant $\tau$ in Theorem 1 of \cite{Shen1994} is $\frac{1-\max\{v,q\}}{2}-\frac{\log\log n}{2\log n}$. According to Lemma \ref{lem:appro} and Condition \ref{assume2}, we have  $d(\xi_{0n},\xi_0)=O(n^{-\min\{\omega\nu,\kappa q\}})$ and
$\left|g_{0n}(r_{\theta_{0n}})-g_{0}(r_{\theta_{0}})\right|=O(n^{-\min\{\omega\nu,\kappa q\}}). $
Based on \eqref{M function}  and the property that $\phi(x)\leq x^2$  when $x\leq 1/2,$ it follows that
\begin{equation*}
	\begin{split}
	K(\xi_{0n},\xi_{0}) \equiv M(\xi_{0})-M(\xi_{0n})
	\leq P\left(\Delta \left\{g_{0n}(r_{\theta_{0n}})-g_{0}(r_{\theta_{0}})\right\}^2\right)
	=O(n^{-2\min\{\omega\nu,\kappa q\}}),
	\end{split}
\end{equation*}
which leads to $K^{1/2\alpha}(\xi_{0n},\xi_{0})= O(n^{-\min\{\omega\nu,\kappa q\}}).$   Based on Case 3 on page 591 of \cite{Shen1994},  since $c_n$ grows with $n$ slowly enough, $A_1$ and $A_2$ in the conditions C1 and C2 can be taken to be independent of $n$.  Consequently,  from Theorem 1 of \cite{Shen1994},   we derive the convergence rate of $\|\hat{g}_n(r_{\hat{\theta}_n})-g_0(r_{\theta_{0}})\|_{\mathcal{A}}$, given as
\begin{equation*}
	\|\hat{g}_n(r_{\hat{\theta}_n})-g_0(r_{\theta_{0}})\|_\mathcal{A}=O_{p}(n^{-c}),
\end{equation*}
where $c=\min\big\{  \omega \nu, \kappa q, (1-\max\{\nu,q\})/2  \big\}.$
\end{proof}

\subsection{Proof of Theorem 2}
\begin{proof}
In the first step, we will derive the consistency of $\hat{\xi}_n$. Based on the result of Theorem 1, it is reasonable to consider $c_n$ as a fixed but sufficiently large value such that $\xi_{0n}$ is included in $\Xi_n$ and this is   what we consider in the following proof.  Clearly,  $\|\hat{g}_n\|_{\infty}=O_p(1)$ and $\|\hat{\beta}_n\|=O_p(1)$.   According to  Lemma 11 in \cite{Stone1985}, $\int_{a}^{b}|\ddot{\hat{g}}_n(s)|^2ds=O_p(1)$.    Consider a class of functions $\mathcal{G}(M)\equiv\{g\in\mathcal{G}^{\kappa}: \|g\|_{\infty}\leq M, \text{ and } \int_{a}^{b}|\ddot{g}(s)|^2ds\leq M^2\}$.   Lemmas 4 and 5 of \cite{Kuchibhotla2020} show that for every $g\in\mathcal{G}(M)$, we have
\begin{equation*}
	|\dot{g}(t)-\dot{g}(s)|\leq M|t-s|^{1/2} \text{ and  }\|\dot{g}\|_{\infty}\lesssim M^2.
\end{equation*}
Therefore, it follows by  Arzel$\grave{a}$-Ascoli theorem that both $\mathcal{G}(M)$ and $\{\dot{g}: g\in\mathcal{G}(M)\}$ have compact closure with respect to  the supremum norm. This implies that  every sequence $\{g_{n}\}_{n\geq 1}$ in $\mathcal{G}(M)$ has a subsequence  such that both $\{\dot{g}_{n_{k}}\}_{k\geq 1}$ and $\{g_{n_{k}}\}_{k\geq 1}$ converges uniformly on $[a,b]$. Let $L_C: L_{2}[0,1] \rightarrow L_{2}[0,1]$ denote an integral operator defined by
\begin{equation*}
L_{C}(\beta)(\cdot)=\int_{0}^{1}C(s,\cdot)\beta(s)ds.
\end{equation*}
The  operator $L_{C}$ is an Hilbert-Schmidt operator, and hence it is a compact operator. Since $L_{2}[0,1]$ is a separable Hilbert space,  every bounded sequence $\{\beta_n\}_{n\geq 1}$ has a weakly convergent subsequence $\{\beta_{n_{k}}\}_{k\geq 1}$, which leads to that $\{L_{C}(\beta_{n_{k}})\}_{k\geq 1}$ is strongly convergent.

Suppose that  we have a sequence $\{(\theta_n, g_{n})\}_{n\geq 1}$, where $\theta_n=(\alpha_n, \beta_n)$.  It satisfies $\|g_n(r_{\theta_n})-g_0(r_{\theta_0})\|_{\mathcal{A}}\rightarrow 0$, $\{g_n\}_{n\geq 1}\in \mathcal{G}(M)$ for some $M$ and $\{\beta_n\}_{n\geq 1}$ is bounded. The above statements indicate  that every subsequence of $\{(\theta_n, g_{n})\}_{n\geq 1}$ has  a subsequence $\{( \theta_{n_{k}}, g_{n_k})\}_{k\geq 1}$ such that $\|g_{n_{k}}-g^{\star}\|_{\infty}+\|\dot{g}_{n_{k}}-\dot{g}^{\star}\|_{\infty}\rightarrow 0$, $|\alpha_{n_{k}}-\alpha^{\star}|\rightarrow 0$ and $\|\beta_{n_{k}}-\beta^{\star}\|_{C}\rightarrow 0,$ for some $g^{\star}, \alpha^{\star}$ and $\beta^{\star}$. Let $\theta^{\star}=(\alpha^{\star},\beta^{\star}).$ From    direct calculation, $( \theta^{\star}, g^{\star})$ satisfies  $\|g^{\star}(r_{\theta^{\star}})-g_0(r_{\theta_0})\|_{\mathcal{A}}=0$. Hence, it  follows by Proposition \ref{pro:identifiability} that when $I(\alpha_0)$ is nonsingular,     $|\alpha^{\star}-\alpha_0|=0$,  $\|\beta^{\star}-\beta_{0}\|_{C}=0,$ and $g^{\star}=g_{0}.$    Based on the result of Theorem 1, we derive that $\|\hat{g}_n-g_0\|_{\infty}+\|\dot{\hat{g}}_{n}-\dot{g}_0\|_{\infty}=o_p(1)$, $|\hat{\alpha}_{n}-\alpha_0|=o_p(1)$ and $\|\hat{\beta}_{n}-\beta_0\|_{C}=o_p(1).$
\medskip

In the next step, we investigate the convergence rate of $\hat{\xi}_n$. According to  Condition \ref{assume6},   it follows that
\begin{equation*}
	\begin{split}
		&\Big| P\Big[\int_{a}^{b}I(r_{\theta_{0}}\geq t)\exp\{g_{0}(t)\} \dot{g}_0(t)\mu(U,\theta-\theta_{0})\{g(t)-g_{0}(t)\}dt\Big]\Big|^2
		\\&=\Big| P\Big[\int_{a}^{b}I(r_{\theta_{0}}\geq t)\exp\{g_{0}(t)\} \dot{g}_0(t)P\left[\mu(U,\theta-\theta_{0})| r_{\theta_{0}}\right]\{g(t)-g_{0}(t)\}dt\Big]\Big|^2
		\\&\leq (1-\eta)\Big| P\Big[\int_{a}^{b}I(r_{\theta_{0}}\geq t)\exp\{g_{0}(t)\} |\dot{g}_0(t)|P\left[\mu(U,\theta-\theta_{0})^2| r_{\theta_{0}}\right]^{1/2}|g(t)-g_{0}(t)|dt\Big]\Big|^2
		\\&\leq (1-\eta)\cdot P\Big[\Delta\dot{g}_0^2(r_{\theta_{0}})\mu(U,\theta-\theta_{0})^2\Big]\cdot P\Big[ \Delta \big\{g(r_{\theta_{0}})-g_{0}(r_{\theta_{0}})\big\}^2\Big].
	\end{split}
\end{equation*}
Let $B(\xi)=\dot{g}_{0}(r_{\theta_{0}})\mu(U,\theta-\theta_{0})+g(r_{\theta_{0}})-g_{0} (r_{\theta_{0}}). $ Based on the above result, we have
\begin{equation*}
	\begin{split}
		P\left[\Delta B(\xi)^2\right]&=P\Big[\Delta \dot{g}_{0}^2(r_{\theta_{0}})\mu(U,\theta-\theta_{0})^2\Big]+P\Big[\Delta \left\{g(r_{\theta_{0}})-g_{0} (r_{\theta_{0}})\right\}^2\Big]
		\\&\quad+2P\Big[\Delta\dot{g}_{0}(r_{\theta_{0}})\mu(U,\theta-\theta_{0})\big\{g(r_{\theta_{0}})-g_{0} (r_{\theta_{0}})\big\}\Big]
		\\&\geq  P\Big[\Delta \dot{g}_{0}^2(r_{\theta_{0}})\mu(U,\theta-\theta_{0})^2\Big]+P\Big[\Delta \left\{g(r_{\theta_{0}})-g_{0} (r_{\theta_{0}})\right\}^2\Big]
		\\&\quad-2\sqrt{1-\eta}\cdot P\Big[\Delta\dot{g}_0^2(r_{\theta_{0}})\mu(U,\theta-\theta_{0})^2\Big]^{1/2}\cdot P\Big[ \Delta \big\{g(r_{\theta_{0}})-g_{0}(r_{\theta_{0}})\big\}^2\Big]^{1/2}
		\\&\geq (1-\sqrt{1-\eta}) \Big(P\Big[\Delta \dot{g}_{0}^2(r_{\theta_{0}})\mu(U,\theta-\theta_{0})^2\Big]+P\Big[\Delta \big\{g(r_{\theta_{0}})-g_{0} (r_{\theta_{0}})\big\}^2\Big]\Big)
		\\&\succsim  d(\xi, \xi_{0})^2.
	\end{split}
\end{equation*}
Let  $C(\xi)=g(r_{\theta})-g(r_{\theta_{0}})-\dot{g}_{0}(r_{\theta_{0}})\mu(U,\theta-\theta_{0})$. The calculation shows that
\begin{equation*}
	\begin{split}
	P\left[\Delta C(\xi)^2\right]&=P\Big[\Delta\big\{g(r_{\theta})-g(r_{\theta_{0}})-\dot{g}_{0}(r_{\theta_{0}})\mu(U,\theta-\theta_{0})\big\}^2\Big]
	\\&=P\Big[\Delta\big\{\left\{ \dot{g}(r_{\tilde{\theta}})-\dot{g}(r_{\theta_{0}})\right\}\mu(U,\theta-\theta_{0})+\left\{\dot{g}(r_{\theta_{0}})-\dot{g}_{0}(r_{\theta_{0}})\right\}\mu(U,\theta-\theta_{0})\big\}^2\Big]
	\\&\leq 2P\Big[\Delta  \int_a^{b}\ddot{g}(s)^2ds|\mu(U,\theta-\theta_{0})|^3+\Delta\left\{\dot{g}(r_{\theta_{0}})-\dot{g}_{0}(r_{\theta_{0}})\right\}^2\mu(U,\theta-\theta_{0})^2  \Big].
	\end{split}
\end{equation*}
Since
$g(r_{\theta})-g_{0} (r_{\theta_{0}} )=B(\xi)+C(\xi),$  it follows that
\begin{equation*}
	\begin{split}
	\|g(r_{\theta})-g_0(r_{\theta_{0}})\|_\mathcal{A}^2=
	P\big(\Delta\big\{ g(r_{\theta})-g_{0} (r_{\theta_{0}} )  \big\}^2\big)&\geq \frac{1}{2}P\big[\Delta B(\xi)^2\big]-P\big[\Delta C(\xi)^2\big].
	\end{split}
\end{equation*}
The result  that we derive  in the first step  indicates that $P[\Delta C(\hat{\xi}_n)^2]=o_p(1)P[\mu(U, \hat{\theta}_n-\theta_0)^2]$, which leads to
\begin{equation*}
	d(\hat{\xi}_n, \xi_{0})^2\lesssim P\big[\Delta B(\hat{\xi}_n)^2\big]\leq O_p(n^{-2c})+o_p(1)P[\mu(U, \hat{\theta}_n-\theta_0)^2].
\end{equation*}
Therefore, we derive
\begin{equation*}
	d(\hat{\xi}_n, \xi_{0})=O_p(n^{-c}),
\end{equation*}
where $c=\min\big\{  \omega \nu, \kappa q, (1-\max\{\nu,q\})/2  \big\}.$  In addition, by \eqref{dis}, when $I(\alpha_0)$ is nonsingular,  we obtain  $	 |\hat{\alpha}_n-\alpha_0|=O_p(n^{-c})$, which in turn implies that $\|\hat{\beta}_n-\beta_0\|_{C}=O_p(n^{-c})$. Therefore, we show that
\begin{equation*}
	|\hat{\alpha}_n-\alpha_0|+\|\hat{\beta}_n-\beta_0\|_{C}+\|\hat{g}_n-g_0\|_{\mathcal{G}}=O_p(n^{-c}).
\end{equation*}

\end{proof}

 \subsection{Proof of Theorem 3}
\begin{proof}
Motivated by the idea of   Theorem 1 in  \cite{Zhao2017}, we need to verify  the following conditions  and then  derive  the asymptotic normality by selecting a specific    direction $\bh^{\star}$. The conditions are
\begin{itemize}
	\item [(i)]$\mathbb{G}_{n}\left\{\dot{m}(\hat{\xi}_{n};\mathcal{W})[\bh]-\dot{m}(\xi_{0};\mathcal{W})[\bh]\right\}
	=o_p(1),$
	\item [(ii)]$P\dot{m}(\xi_{0};\mathcal{W})[\bh]=0$ and $\mathbb{P}_{n}\dot{m}(\hat{\xi}_{n};\mathcal{W})[\bh]=o_p(n^{-1/2})$,
	\item [(iii)]$P\dot{m}(\hat{\xi}_{n};\mathcal{W})[\bh]-P\dot{m}( \xi_{0};\mathcal{W})[\bh]
	-P\ddot{m}(\xi_{0};\mathcal{W})[\bh][\hat{\xi}_n-\xi_0]=o_p(n^{-1/2}).$
\end{itemize}

  In the first step, we consider general direction $\bh=(h_1,h_2,h_3)\in\mathcal{H}$, where  $h_3\in\mathcal{G}^{\kappa-1}$.    For every $\dot{m}(\xi;\mathcal{W})[\bh]-\dot{m}(\xi_{0};\mathcal{W})[\bh]\in \mathcal{F}_{n}(\bh,\eta)$, where   $\mathcal{F}_{n}(\bh,\eta)$ is defined in Lemma \ref{lem4},    Conditions \ref{assume2} and \ref{assume4} imply that
  \begingroup
  \allowdisplaybreaks
 	\begin{align*}
		&	\Big|
			\dot{m}(\xi;\mathcal{W})[\bh]-\dot{m}(\xi_{0};\mathcal{W})[\bh]
			\Big|
			\\&=\Big| \int_{a}^{b}I(r_{\theta_{0}}\geq t) \Big[\exp\{g(t_{\theta})\}\dot{g}(t_{\theta})-\exp\{g_0(t)\}\dot{g_0}(t)\Big]\mu(U,\delta)
			dt
			\\&\quad -\Delta\big[ \dot{g}(r_{\theta})- \dot{g_0}(r_{\theta_0})\big] 	\mu(U,\delta)
			+\Delta\big [h_3(r_{\theta}) -h_3(r_{\theta_0})\big]
			\\&\quad  -\int_{a}^{b}I(r_{\theta_{0}}\geq t)  \Big[h_3(t_{\theta})\exp \{g(t_{\theta})\}-h_3(t)\exp \{g_0(t)\}\Big]dt\Big|
			\\&\lesssim \Big| \int_{a}^{b}I(r_{\theta_{0}}\geq t) \Big[\exp\{g(t_{\theta})\}\dot{g}(t_{\theta})-\exp\{g_0(t)\}\dot{g_0}(t)\Big] dt  \Big|
			\\&\quad+ \Big|\Delta\big[ \dot{g}(r_{\theta})- \dot{g_0}(r_{\theta_0})\big] \Big|+ \Big|\Delta\big [h_3(r_{\theta}) -h_3(r_{\theta_0})\big]\Big|
			\\&\quad+ \Big|\int_{a}^{b}I(r_{\theta_{0}}\geq t)  \Big[h_3(t_{\theta})\exp \{g(t_{\theta})\}-h_3(t)\exp \{g_0(t)\}\Big]dt\Big|
			\\&=I_{1n}+I_{2n}+I_{3n}+I_{4n},
	\end{align*}
  \endgroup
where $\delta=(h_1,h_2).$ Since for any  $g\in\mathcal{G}_{n}^{\kappa},$
$\|g\|_{\infty}\leq c_n$, from the mean value theorem and Condition   \ref{assume4}, we have
\begin{equation*}
	\begin{split}
		I_{1n}&= \Big| \int_{a}^{b}I(r_{\theta_{0}}\geq t)\Big[             \exp\{g(t_{\theta})\}\big\{\dot{g}(t_{\theta})-\dot{g}_0(t_{\theta})\big\}
		+\dot{g}_{0}(t_{\theta})\big\{\exp\{g(t_{\theta})\}-\exp\{g_{0}(t_{\theta})\}\big\}
		 \\&\qquad  +\dot{g}_{0}(t_{\theta}) \exp\{g_{0}(t_{\theta})\}-\dot{g}_{0}(t) \exp\{g_{0}(t)\}\Big]dt\;\Big|
		 \\& \lesssim  \int_{a}^{b}I(r_{\theta_{0}}\geq t)\Big[\exp\{c_n\}\big|\dot{g}(t_{\theta})-\dot{g}_0(t_{\theta})\big|
		 + \big|\exp\{g(t_{\theta})\}-\exp\{g_{0}(t_{\theta})\}\big|\Big]dt
		 \\&\qquad +\big| \mu(U,\theta-\theta_{0})\big|.
	\end{split}
\end{equation*}
Then, it follows by Condition   \ref{assume2} and \ref{assume4} that
\begin{equation*}
P\{I_{1n}\}^{2}\lesssim\exp\{2c_n\}\eta^2\; \text{  and  }\;I_{1n}\lesssim c_n\exp\{c_n\}.
\end{equation*}
 For $I_{4n}$, with a similar argument, we have
\begin{equation*}
\begin{split}
 	I_{4n}&= \Big| \int_{a}^{b}I(r_{\theta_{0}}\geq t)\Big[\exp\{g(t_{\theta})\}\big\{h_3(t_{\theta})-h_3(t)\big\}
 +h_3(t)\big\{\exp\{g(t_{\theta})\}-\exp\{g(t)\}\big\}
 \\&\qquad  +h_3(t) \big\{\exp\{g(t)\}- \exp\{g_{0}(t)\}\big\}\Big]dt\;\Big|
 \\& \lesssim  \exp\{c_n\}\|\dot{h}_3\|_{\infty}\big| \mu(U,\theta-\theta_{0})\big|+
 \|h_3\|_{\infty}c_{n}\exp\{c_n\}\big| \mu(U,\theta-\theta_{0})\big|
\\&\quad+ \int_{a}^{b}I(r_{\theta_{0}}\geq t) \|h_3\|_{\infty}\big|\exp\{g(t)\}- \exp\{g_{0}(t)\}\big|dt.
\end{split}
\end{equation*}
  Condition \ref{assume2} leads to
\begin{equation*}
P\{I_{4n}\}^2\lesssim c_n^2\exp\{2c_n\}\eta^2\; \text{  and  }\;I_{4n}\lesssim c_n\exp\{c_n\}.
\end{equation*}
For $I_{2n}$ and $I_{3n}$,  according to Conditions \ref{assume2}--\ref{assume4},   it follows that
\[
P\{I_{2n}+I_{3n}\}^2\lesssim \eta^2\; \text{  and  }\;  I_{2n}+I_{3n}\lesssim \exp\{c_{n}\}.
\]
Therefore, we show that
\begin{equation*}
	\begin{split}
		\Big|
		\dot{m}(\xi;\mathcal{W})[\bh]-\dot{m}(\xi_{0};\mathcal{W})[\bh]
		\Big|\lesssim c_{n}\exp\{c_{n}\}
	\end{split}
\end{equation*}
and
\begin{equation*}
	P\big\{
	\dot{m}(\xi;\mathcal{W})[\bh]-\dot{m}(\xi_{0};\mathcal{W})[\bh]
	\big\}^2\lesssim c_n^2\exp\{2c_n\}\eta^2\equiv\lambda_n^2.
\end{equation*}
  According to  Lemma 11 in \cite{Stone1985}, for $c=\min\big\{  \omega \nu, \kappa q, (1-\max\{\nu,q\})/2  \big\}$,  we derive $\|\dot{\hat{g}}_n-\dot{g}_0\|_{\mathcal{G}}=O_p(n^{-c+q})$.
Pick $\eta$ as $\eta_{n}=O(n^{ -c+q })$.
According to the result of Lemma \ref{lem4} that the $\varepsilon$-bracketing numbers for the class $\mathcal{F}_{n}(\bh,\eta_{n})$ is bounded by $(\eta_{n}/\varepsilon)^{c_3m_{n}^{\omega}+c_4s_{n}^{\kappa}+p}$ for  constants $c_3,c_4>0, $  it follows that the bracketing integral satisfies
\begin{equation*}
	\begin{split}
    J_{[\;]}(\lambda_n, \mathcal{F}_{n}(\bh,\eta_{n}), L_{2}(P))&=  \int^{\lambda_n}_{0}\sqrt{1+\log N_{[\;]}(\varepsilon,\mathcal{F}_{n}(\bh,\eta_{n}),  L_{2}(P))} d\varepsilon
    \\&\leq \int^{\lambda_n}_{0}\sqrt{1+\log N_{[\;]}(\varepsilon,\mathcal{F}_{n}(\bh,\eta_{n}), \|\cdot\|_{\infty})} d\varepsilon
    \\&\lesssim \lambda_n\sqrt{c_3m_{n}^{\omega}+c_4s_{n}^{\kappa}+p}.
	\end{split}
\end{equation*}
Then, by the maximal  inequality in Lemma 3.4.2 of \cite{Van1996}, we have
\begin{equation*}
	\begin{split}
		E_{P}\|\mathbb{G}_{n}\|_{\mathcal{F}_{n}(\bh,\eta_n)}
		&\lesssim \exp\{2c_n\}\Big[\eta_{n}\sqrt{c_3m_{n}^{\omega}+c_4s_{n}^{\kappa}+p}+\frac{c_3m_{n}^{\omega}+c_4s_{n}^{\kappa}+p}{\sqrt{n}}\Big]
		\\&=\exp\{2c_n\}\cdot O\left(n^{\frac{\max\{\nu, q\}}{2}-c+q}\right)
		\\&=o(1),
	\end{split}
\end{equation*}
 where the last equality  holds since   $c-q>\frac{1}{4}$, $\max\{\nu,q\}<\frac{1}{2}$, and $c_n$ grows with $n$ slowly enough. Therefore, for a constant $C $,   Markov's inequality leads to
\begin{equation*}
	\begin{split}
	  \sup_{\{\xi\in\Xi_{n}:\;d(\xi,\xi_{0})\leq Cn^{-c}\}}\mathbb{G}_{n}\left\{\dot{m}(\xi;\mathcal{W})[\bh]-\dot{m}(\xi_{0};\mathcal{W})[\bh]\right\}
=o_p(1),
	\end{split}
\end{equation*}
 which  demonstrates that
\begin{equation*}\label{conA1}
\mathbb{G}_{n}\left\{\dot{m}(\hat{\xi}_{n};\mathcal{W})[\bh]-\dot{m}(\xi_{0};\mathcal{W})[\bh]\right\}
=o_p(1).
\end{equation*}
Thus, condition (i) holds.

Based on the model assumptions, $
	P\dot{m}(\xi_{0};\mathcal{W})[\bh]=0$ holds automatically.
Next, we   investigate
$\mathbb{P}_{n}\dot{m}(\hat{\xi}_{n};\mathcal{W})[\bh]$ and show it is $o_p(n^{-1/2}).$  From Lemma \ref{lem:appro}, there exists $h_{2n}\in\mathcal{F}_{n}^{\omega}$ and $h_{3n}\in\mathcal{G}_{n}^{\kappa} $ such that  $\|h_{2n}-h_{2}\|_{\infty}=O(n^{-\omega\nu})$ and $\|h_{3n}-h_{3}\|_{\infty}=O(n^{-(\kappa-1)q})$.  Define $\bh_{n}=(h_1,h_{2n}, h_{3n})$.  Since $\hat{\xi}_{n}$ maximizes the log-likelihood function $l_n$ on the sieve space $\Xi_{n}$, it follows that  $\mathbb{P}_{n}\dot{m}(\hat{\xi}_{n};\mathcal{W})[\bh_{n}]=0$.  Hence, it  suffices to study $\mathbb{P}_{n}\dot{m}(\hat{\xi}_{n};\mathcal{W})[\bh-\bh_n]$, which is decomposed as
 \begin{equation*}
 	\begin{split}
 \mathbb{P}_{n}\dot{m}(\hat{\xi}_{n};\mathcal{W})[\bh-\bh_n]&=B_{1n}+B_{2n},
  	\end{split}
\end{equation*}
where
\begin{equation*}
	\begin{split}
&B_{1n}=\big\{\mathbb{P}_{n}-P\big\}\dot{m}(\hat{\xi}_{n};\mathcal{W})[\bh-\bh_n]\quad\text{  and  }
\\&B_{2n}=P
 \dot{m}(\hat{\xi}_{n};\mathcal{W})[\bh-\bh_n]-P\dot{m}( \xi_{0};\mathcal{W})[\bh-\bh_n].
 	\end{split}
 \end{equation*}
 We will show that both $B_{1n}$ and $B_{2n}$ are  $o_p(n^{-1/2})$.

To investigate  $B_{1n}$, we  consider the class of functions $\mathcal{L}_{n}(\bh,\eta)$  defined in Lemma \ref{lem5}.  For any $\dot{m}(\xi;\mathcal{W})[\bh-\tilde{\bh}]\in\mathcal{L}_{n}(\bh,\eta)$,   Condition \ref{assume2} implies that
\begin{equation*}
	\begin{split}
		\Big|\dot{m}(\xi;\mathcal{W})[\bh-\tilde{\bh}]\Big|&=\Big|\Big\{\int_{a}^{b}I(r_{\theta_{0}}\geq t )  \exp\{g(t_{\theta})\}\dot{g}(t_{\theta})
		dt
		-\Delta \dot{g}(r_{\theta})\Big\}	\int_{0}^{1}\{h_2-\tilde{h}_2\}(s)Z(s)ds
		\\&\quad+\Delta \{h_3-\tilde{h}_3\}(r_{\theta})    -\int_{a}^{b}I(r_{\theta_{0}}\geq t)   \{h_3-\tilde{h}_3\}(t_{\theta})\exp \{g(t_{\theta})\}dt\Big|
		\\&\lesssim\|h_2-\tilde{h}_2\|_{\infty}  \Big| \int_{a}^{b}I(r_{\theta_{0}}\geq t )  \exp\{g(t_{\theta})\}\dot{g}(t_{\theta})
		dt
		-\Delta \dot{g}(r_{\theta})\Big|
		\\&\quad+ \|h_3-\tilde{h}_3\|_{\infty}+\|h_3-\tilde{h}_3\|_{\infty}\Big|\int_{a}^{b}I(r_{\theta_{0}}\geq t)   \exp \{g(t_{\theta})\}dt\Big|
		\\&\lesssim  c_n\exp\{c_n\}\eta\equiv\lambda_n,
	\end{split}
\end{equation*}
which  also demonstrates that $P\{\dot{m}(\xi;\mathcal{W})[\bh-\tilde{\bh}]\}^2\lesssim \lambda_n^2.$
Pick $\eta$ as $\eta_{n}=O(n^{-c+q}).$  According to the result of Lemma \ref{lem5}   that the $\varepsilon$-bracketing number associated with  the supremum  norm for the class of functions $\mathcal{L}_{n}(\bh,\eta_n)$   is bounded by $ (\eta_n/\varepsilon)^{c_5m_{n}^{\omega}+c_6s_{n}^{\kappa}+p} $  for constants $c_5, c_6>0$, the bracketing integral  satisfies
 \begin{equation*}
 	\begin{split}
 	 	J_{[\;]}(\lambda_n,\mathcal{L}_{n}(\bh,\eta_n),L_2(P))
 	 	 &\leq \int^{\lambda_n}_{0}\sqrt{1+\log N_{[\;]}(\varepsilon,\mathcal{L}_{n}(\bh,\eta_n), \|\cdot\|_{\infty})} d\varepsilon
 		\\&\lesssim \lambda_n\sqrt{c_5m_{n}^{\omega}+c_6s_{n}^{\kappa}+p}.
 	\end{split}
 \end{equation*}
With a similar argument as before,  the maximal  inequality in Lemma 3.4.2 of \cite{Van1996} leads to
\begin{equation*}
	\begin{split}
		E_{P}\|\mathbb{G}_{n}\|_{\mathcal{L}_{n}(\bh,\eta_n)}=o(1).
	\end{split}
\end{equation*}
Based on $\|\bh_{n}-\bh\|_{\infty}=O(n^{-\min\{\omega\nu, (\kappa-1)q\}})$, it follows that $\bh_{n}\in\mathcal{H}_{n}(\bh,\eta_{n})$. Therefore, by Markov's inequality, we have
\begin{equation*}
	\begin{split}
	B_{1n}=\big\{\mathbb{P}_{n}-P\big\}\dot{m}(\hat{\xi}_{n};\mathcal{W})[\bh-\bh_n]=o_p(n^{-1/2}).
		\end{split}
\end{equation*}

To study $B_{2n}$,    we  apply the mean value theorem and get
\begin{equation}\label{eq:decomp}
	\begin{split}
		&\;	\Big|B_{2n}-P\ddot{m}(\xi_{0};\mathcal{W})[\bh-\bh_n][\hat{\xi}_n-\xi_0]\Big|
		\\&=\Big|P
		\dot{m}(\hat{\xi}_{n};\mathcal{W})[\bh-\bh_n]-P\dot{m}( \xi_{0};\mathcal{W})[\bh-\bh_n]
		-P\ddot{m}(\xi_{0};\mathcal{W})[\bh-\bh_n][\hat{\xi}_n-\xi_0]\Big|
		\\&\leq    \Big|P\ddot{m}(\tilde{\xi}_n;\mathcal{W})[\bh-\bh_n][\hat{\xi}_n-\xi_0]-P\ddot{m}(\xi_{0};\mathcal{W})[\bh-\bh_n][\hat{\xi}_n-\xi_0]\Big|.
	\end{split}
\end{equation}
where $\tilde{\xi}_n$ is between $\xi_{0}$ and $\hat{\xi}_n.$
Denote $\delta_{n}=(0,h_2-h_{2n})$ and $\theta_{n}=(\hat{\alpha}_n-\alpha_{0},\hat{\beta}_{n}-\beta_{0})$.  From Conditions \ref{assume2} and \ref{assume4},   direct calculation yields
  \begingroup
\allowdisplaybreaks
\begin{align*}
&\Big|P\ddot{m}(\xi_{0};\mathcal{W})[\bh-\bh_n][\hat{\xi}_n-\xi_0]\Big|
\\&=\Big|
P\Big[
-\Delta\dot{g}_{0}^2(r_{\theta_{0}})\mu(U,\delta_{n})\mu(U,\theta_{n})+\Delta\dot{g}_{0}(r_{\theta_{0}})\{\hat{g}_n-g_0\}(r_{\theta_{0}})\mu(U,\delta_{n})
\\&\qquad\quad+\Delta\dot{g}_{0}(r_{\theta_{0}})\{h_{3}-h_{3n}\}(r_{\theta_{0}})\mu(U,\theta_{n})-\Delta\{\hat{g}_n-g_0\}(r_{\theta_{0}})\{h_{3}-h_{3n}\}(r_{\theta_{0}}) \Big]
\Big|	
\\&\lesssim \|\dot{g}_{0}^2\|_{\infty}\|\bh_{n}-\bh\|_{\infty}\cdot d(\hat{\xi}_{n},\xi_{0})+\|\dot{g}_{0}\|_{\infty}\|\hat{g}_n-g_0\|_{\mathcal{G}}\cdot \|\bh_{n}-\bh\|_{\infty}
\\&\qquad+\|\dot{g}_{0}\|_{\infty}\|h_{3n}-h_{3}\|_{\mathcal{G}}\cdot d(\hat{\xi}_{n},\xi_{0})
+\|\hat{g}_n-g_0\|_{\mathcal{G}}\cdot\|h_{3n}-h_{3}\|_{\mathcal{G}}
\\&\lesssim\|\bh_{n}-\bh\|_{\infty}\cdot d(\hat{\xi}_{n},\xi_{0})
\\&=O(n^{-\min\{\omega\nu,(\kappa-1)q\}})\cdot O_p(n^{-c})
\\&=o_p(n^{-1/2}),
\end{align*}
\endgroup
 where the last  equality holds since   $\omega\nu>\frac{1}{4}$ and $(\kappa-1)q>\frac{1}{4}$ under the conditions of the theorem. Therefore,  to verify $B_{2n}=o_p(n^{-1/2})$, it  suffices to show that   the right-hand side of the last inequality of \eqref{eq:decomp} is $o_p(n^{-1/2})$.

 Define $Q_n(\xi,\bh)=\big|P\ddot{m}(\xi;\mathcal{W})[\bh][\hat{\xi}_n-\xi_0]-P\ddot{m}(\xi_{0};\mathcal{W})[\bh][\hat{\xi}_n-\xi_0]\big|$, where $\xi$ satisfies  $\|g\|_{\infty} , \|\dot{g}\|_{\infty}, \|\ddot{g}\|_{\infty}\leq c_n$ and $ \|\bh\|_{\infty}, \|\dot{h}_{3}\|_{\infty}$ are   bounded.
 Denote $\delta=(h_1, h_2)$ and  $g_n=\hat{g}_{n}-g_{0}$. Direct calculation yields
\begin{equation*}
\begin{split}
		Q_n(\xi,\bh)&\leq \Big|P\Big[\Delta\mu(U,\delta)\mu(U,\theta_{n})\big\{  \exp\{g(r_{\theta})-g_0(r_{\theta_{0}})\}\{ \ddot{g}+(\dot{g})^2\}(r_{\theta})
                     - \ddot{g} (r_{\theta}) - (\dot{g}_0)^2(r_{\theta_{0}})   \big\}\Big]\bigg|
		\\&\quad +\bigg|P\Big[\Delta\mu(U,\delta) \big\{
		\exp \{g(r_{\theta})-g_0(r_{\theta_0})\} \{ \dot{g}_{n}+\dot{g}g_{n}\}(r_{\theta})	-  \dot{g}_{n}(r_{\theta})
		-  \{\dot{g}_{0}g_{n}\}(r_{\theta_{0}})
		\big\}\Big]\bigg|
		\\&\quad +\bigg|P\Big[\Delta\mu(U,\theta_{n})\big\{
		\exp \{g(r_{\theta})-g_0(r_{\theta_0})\} \{\dot{h}_3+\dot{g}h_3\}(r_{\theta})- \dot{h}_3(r_{\theta})-\{\dot{g}_{0}h_3\}(r_{\theta_{0}})
		\big\}\Big]\bigg|
		\\&\quad+\bigg|P\Big[\Delta\big\{ \exp\left\{g(r_{\theta})-g_0(r_{\theta_0})\right\} h_3(r_{\theta})g_{n}(r_{\theta})-   h_3(r_{\theta_{0}})g_{n}(r_{\theta_{0}})\big\} \Big]\bigg|
		\\&=Q_{1n}+Q_{2n}+Q_{3n}+Q_{4n}.
	\end{split}
\end{equation*}
 For $Q_{1n}$, based on inequality $|\exp(x)-1|\leq |x|\exp(|x|),$ Conditions \ref{assume2} and \ref{assume4} lead to
\begin{equation*}
	\begin{split}
		Q_{1n}&\lesssim   \Big[P\Big\{ \Delta\big(\exp\{g(r_{\theta})-g_0(r_{\theta_{0}})\}-1\big)\ddot{g} (r_{\theta})\Big\}^2
		+P\Big\{\Delta\big(\exp\{g(r_{\theta})-g_0(r_{\theta_{0}})\}-1\big)\dot{g}^2(r_{\theta})\Big\}^2\\&\qquad+
		P\Big\{\Delta\big( \dot{g}^2(r_{\theta})-\dot{g}_0^2(r_{\theta_{0}}) \big)\Big\}^2\;\Big]^{1/2}\cdot d(\hat{\xi}_n,\xi_{0})\cdot \|\bh\|_{\infty}
		\\&\lesssim c_n^2\exp\{c_n\} d(\hat{\xi}_n,\xi_{0}) \|\bh\|_{\infty}  \Big[P\Big\{\Delta\big(g(r_{\theta})-g_{0}(r_{\theta_{0}})\big)\Big\}^2
		+P\Big\{\Delta\big(\dot{g}(r_{\theta})-\dot{g}_{0}(r_{\theta_{0}})\big)\Big\}^2\;
		\Big]^{1/2}
		\\&\lesssim c_n^2\exp\{c_n\}\zeta d(\hat{\xi}_n,\xi_{0}),
	\end{split}
\end{equation*}
where $\zeta=d(\xi,\xi_{0})+\|\dot{g}-\dot{g}_{0}\|_{\mathcal{G}}$.
Similarly, we can show that
\begin{equation*}
	\begin{split}
	&\begin{split}
			Q_{2n}&\lesssim \Big[P\Big\{\Delta
		\big(\exp\{g(r_{\theta})-g_0(r_{\theta_{0}})\}-1\big)\Big\}^2P\Big\{\Delta  \dot{g}_{n}(r_{\theta})\Big\}^2+\Big|
			P\Big\{\Delta   \{\dot{g}g_{n}\}(r_{\theta})
			-  \Delta \{\dot{g}_{0}g_{n}\}(r_{\theta_{0}})  \Big\}\Big|^2
			\\ &\qquad+P\Big\{\Delta
			\big(\exp\{g(r_{\theta})-g_0(r_{\theta_{0}})\}-1\big)\Big\}^2P\Big\{\Delta\{\dot{g}g_{n}\}(r_{\theta})\Big\}^2\;
			  \Big]^{1/2} \cdot \|\bh\|_{\infty}
			 \\&\lesssim  c_{n}\exp\{c_n\}\zeta (d(\hat{\xi}_n,\xi_{0})+\|\dot{\hat{g}}_{n}-\dot{g}_0\|_{\mathcal{G}} )
		\end{split}
	\\&\begin{split}
		Q_{3n}&\lesssim  \Big[P\Big\{\Delta  \big(
	\exp\{g(r_{\theta})-g_0(r_{\theta_{0}})\}-1\big) \dot{h}_3(r_{\theta})\Big\}^2+
		P\Big\{\Delta  \{\dot{g}h_3\}(r_{\theta})
		-  \Delta\{\dot{g}_{0}h_3\}(r_{\theta_{0}}) \Big\}^2
		\\ &\qquad+P\Big\{\Delta 	\big(
		\exp\{g(r_{\theta})-g_0(r_{\theta_{0}})\}-1\big)\{\dot{g}h_3\}(r_{\theta})\Big\}^2\;
		\Big]^{1/2} \cdot d(\hat{\xi}_n,\xi_{0})
		\\&\lesssim c_n^2\exp\{c_n\} \zeta d(\hat{\xi}_n,\xi_{0}),
	\end{split}
\\&\begin{split}
	Q_{4n}&\lesssim \Big[ P\Big\{\Delta \big(\exp\{g(r_{\theta})-g_0(r_{\theta_{0}})\}-1\big)\Big\}^2P\Big\{\Delta h_3(r_{\theta}) g_{n}(r_{\theta})\Big\}^2
	\\&\qquad+\Big|P\Big\{\Delta\big(h_3(r_{\theta})g_{n}(r_{\theta})-h_3(r_{\theta_{0}})g_{n}(r_{\theta_{0}})\big)\Big\}	\Big|^2\;\Big]^{1/2}
\\& \lesssim  \exp\{c_n\}\zeta(d(\hat{\xi}_n,\xi_{0})+\|\dot{\hat{g}}_{n}-\dot{g}_0\|_{\mathcal{G}} ).
\end{split}
	\end{split}
\end{equation*}
Therefore,   it follows that
$$ Q_n(\xi,\bh)\lesssim c_n^2\exp\{c_n\}\zeta( d(\hat{\xi}_n,\xi_{0})+\|\dot{\hat{g}}_{n}-\dot{g}_0\|_{\mathcal{G}} ).$$
Pick $\xi=\tilde{\xi}_n$.  Then, $\zeta=d(\tilde{\xi}_n,\xi_{0})+\|\dot{\tilde{g}}_n-\dot{g}_{0}\|_{\mathcal{G}}=O_p(n^{-c+q})$.   Since    $c_n$ grows with $n$ slowly enough,  it leads to $Q_n(\tilde{\xi}_n,\bh)=o_p(n^{-1/2})$.
Therefore,  we show that
\begin{equation*}\label{conA5}
	P\dot{m}(\hat{\xi}_{n};\mathcal{W})[\bh]-P\dot{m}( \xi_{0};\mathcal{W})[\bh]
	-P\ddot{m}(\xi_{0};\mathcal{W})[\bh][\hat{\xi}_n-\xi_0]=o_p(n^{-1/2}),
\end{equation*}
which verifies condition (iii).
Taking  $\bh$ as $\bh-\bh_n,$ with a similar argument,   it follows that
$Q_n(\tilde{\xi}_n,\bh-\bh_n)=o_p(n^{-1/2}),$
which completes our verification of $B_{2n}=o_p(n^{-1/2})$. Therefore, we show that
\begin{equation*}
\mathbb{P}_{n}\dot{m}(\hat{\xi}_{n};\mathcal{W})[\bh]=o_p(n^{-1/2}),
\end{equation*}
which verifies condition (ii). Combining conditions (i)--(iii), we obtain
\begin{equation}\label{result}
	-\sqrt{n}P\ddot{m}(\xi_{0};\mathcal{W})[\bh][\hat{\xi}_n-\xi_0]=\sqrt{n}\mathbb{P}_n \dot{m}(\xi_{0};\mathcal{W})[\bh]+o_p(1).
\end{equation}

To derive the result of the theorem, we consider a specific direction  $\bh^{\star}$. For each $i=1,\dots,p,$ we  pick
 $\bh_i^{\star}=(e_{i},\hs_{2i},\hs_{3i})$, where  $e_i$ denotes the vector with a $1$ in the $i$-th coordinate and 0's elsewhere, and $\hs_{2i}$ as well as $\hs_{3i}$ are selected such that
$ P\ddot{m}(\xi_{0};\mathcal{W})[\bh_i^{\star}][\bh]=0$  for any $\bh=(0,h_2,h_3)\in\mathcal{H}$.
To find such  $\hs_{2i}$ and $\hs_{3i}$, it  suffices to solve
\begin{equation}\label{pro}
	 \underset{h_2\in\mathcal{H}_\beta,h_3\in\mathcal{H}_g} {\min}P\Big[\Delta\Big\{\dot{g}_{0}(r_{\theta_{0}})e_i^{\top}X-\eta(r_{\theta_{0}};h_2,h_3)\Big\}^2\;\Big],
\end{equation}
where $\eta(t;h_2,h_3)= \dot{g}_{0}(t)\int_{0}^{1}h_2(s)Z(s)ds-h_3(t).$  
Let $$
	\bm{\ell}^{\star}_{\xi_{0}}(\mathcal{W})=\left( \dot{m}(\xi_{0};\mathcal{W})[\bh_1^{\star}],\dots, \dot{m}(\xi_{0};\mathcal{W})[\bh_{p}^{\star}]  \right)^{\top}, $$ and recall that $M(t)=\Delta I(r_{\theta_{0}}\leq t)-\int^{t}_{-\infty}I(r_{\theta_{0}}\geq u)\lambda_{0}(u)du$ is the counting process martingale. Based on Proposition \ref{pro:score}, it follows that
	\begin{equation*}
		\bm{\ell}^{\star}_{\xi_{0}}(\mathcal{W})=\int_{a}^{b}-\dot{g}_{0}(t)X+\eta(t; b^{\star},\phi^{\star})dM(t),
	\end{equation*}
 is the efficient score for $\alpha_0$.
Let $\textbf{A}=(A_{ij})$ be the $p\times p$ matrix with $A_{ij}= -P\ddot{m}(\xi_{0};\mathcal{W})[\bh_{i}^{\star}][\bh_{j}^{\star}]=-P\ddot{m}(\xi_{0};\mathcal{W})[\bh_{i}^{\star}][(e_j,0,0)].$ Then, we have
\begin{equation*}
\textbf{A}=(A_{ij})=P[\bm{\ell}^{\star}_{\xi_{0}}(\mathcal{W})^{\otimes 2}],
\end{equation*}
where $x^{\otimes 2}=xx^{\top}$ for any vector $x\in\mathbb{R}^{p}.$
Hence, $\textbf{A}=I(\alpha_0)$, which is the information for estimation of $\alpha_0$.
Combining the result with \eqref{result}, we show that
\begin{equation*}
\sqrt{n}(\hat{\alpha}_{n}-\alpha_{0})=I(\alpha_0)^{-1}\sqrt{n}\mathbb{P}_{n}	 \bm{\ell}^{\star}_{\xi_{0}}(\mathcal{W})+o_p(1)\stackrel{D}{\longrightarrow} N(0,  \Sigma ),
\end{equation*}
where $ \Sigma =I(\alpha_0)^{-1}.$ Hence, the proof is completed.
\end{proof}

 \section{Proofs of Auxiliary Lemmas}\label{app-B}
 \subsection{Proof of Lemma \ref{lem3}}\label{pf:lem3}
\begin{proof}
	Denote the ceiling of $x$ by $\lceil x\rceil$. For the spaces $\mathcal{F}_{n}^{\omega}$ and $\mathcal{G}_{n}^{\kappa},$ according to the calculation of \citet{Shen1994} on page 597,   for any $\varepsilon>0$, there exists  sets of brackets
	\begin{equation}\label{bracket}
	\left\{ [\beta_{i}^{L}, \beta_{i}^{U}]: i=1,2, \dots,\left\lceil(1 / \varepsilon)^{c_{1} m_{n}^{\omega}}\right\rceil\right\}\text{ and }\left\{ [g_{s}^{L}, g_{s}^{U} ]:s=1,2, \dots,\left\lceil(1 / \varepsilon)^{c_{2} s_{n}^{\kappa}}\right\rceil\right\},
	\end{equation}
	such that for any $\beta \in \mathcal{F}_{n}^{\omega}$ and  $g \in \mathcal{G}_{n}^{\kappa}$, one has
	\begin{equation*}
	\beta_{i}^{L}(t) \leq \beta(t) \leq \beta_{i}^{U}(t),\;t\in[0,1] \;\text{ and }\; g_{s}^{L}(t) \leq g(t) \leq g_{s}^{U}(t),\; t \in[a, b]
	\end{equation*}
 for some $i\in\{1,\dots,\lceil(1 / \varepsilon)^{c_{1} m_{n}^{\omega}}\rceil\} $  and $ s\in\{1,\dots,\lceil(1 / \varepsilon)^{c_{2} s_{n}^{\kappa}}\rceil\}.$  In addition,   the brackets given  in \eqref{bracket} satisfy
 \begin{equation*}
 	 \left\|\beta_{i}^{U}-\beta_{i}^{L}\right\|_{\infty} \leq \varepsilon\;\text{ and }\;\left\|g_{s}^{U}-g_{s}^{L}\right\|_{\infty} \leq \varepsilon.
 \end{equation*}
	Under Condition \ref{assume1} that   $\mathcal{B} \subseteq \mathbb{R}^{p}$ is compact, $\mathcal{B}$ can be covered by $\left\lceil c_{3}(1 / \varepsilon)^{p}\right\rceil$ balls with radius $\varepsilon$ for a constant $c_3$. Hence,  for any $\alpha \in \mathcal{B}$, there exists $\ell\in\{1,\dots \lceil c_{3}(1 / \varepsilon)^{p}\rceil\}$ such that $\left|\alpha-\alpha_{\ell}\right| \leq  \varepsilon$. Based on condition \ref{assume2}, we have
	\begin{equation*}
		\left|X^{\top}\alpha-X^{\top}\alpha_{\ell}\right| \leq c_{4}\varepsilon \text{ and }  \int_{0}^{1}\mid \beta(s)Z(s)-\beta_{i}^{U}(s) Z(s)\mid ds\leq c_5\varepsilon
	\end{equation*}
     for constants $c_4, c_5>0$.  Let $\theta_{\ell,i}=(\alpha_{\ell}, \beta_{i}^{U})$ and $C=c_4+c_5$. It follows that  for any $t$,
      \begin{equation*}
      t_{\theta_{\ell,i}}-C\varepsilon	\leq t_{\theta}\leq  t_{\theta_{\ell,i}}+C\varepsilon,
     \end{equation*}
 where $t_{\theta}=t-\mu(U,\theta-\theta_{0}).$  Assume $g_{s}^{L}\left(t_{\theta_{\ell,i}}+c_{\ell,i,s,t}^{L}\varepsilon\right)$ and $g_{s}^{U}\left(t_{\theta_{\ell,i}}+c_{\ell,i,s,t}^{U}\varepsilon\right)$
	are the minimum and maximum values of $g_{s}^{L}$ and $g_{s}^{U}$ within the interval $[   t_{\theta_{\ell,i}}-C\varepsilon, t_{\theta_{\ell,i}}+C\varepsilon]$, where $c_{\ell,i,s,t}^{L}$ and $c_{\ell,i,s,t}^{U}$ are two constants that satisfy $\left| c_{\ell,i,s,t}^{L}\right|,\left|c_{\ell,i,s,t}^{U}\right| \leq C $ and  only depend on $g_{s}^{L}, g_{s}^{U}$ and  $t$.  It follows that
\begin{equation*}
		\begin{split}
		g_{s}^{L}\left(t_{\theta_{\ell,i}}+c_{\ell,i,s,t}^{L}\varepsilon\right) \leq g_{s}^{L}\left(t_{\theta}\right)
		 \leq g\left(t_{\theta}\right) \leq g_{s}^{U}\left(t_{\theta}\right)  \leq g_{s}^{U}\left(t_{\theta_{\ell,i}}+c_{\ell,i,s,t}^{U}\varepsilon\right).
	\end{split}
\end{equation*}
	Therefore,  we can construct a set of brackets
\begin{equation*}
	\begin{split}
		\bigg\{\left[m_{\ell,i,s}^{L}(\mathcal{W}), m_{\ell,i,s}^{U}(\mathcal{W})\right]&: \ell=1,\dots \lceil c_{3}(1 / \varepsilon)^{p}\rceil;  i=1, \dots,\lceil(1 / \varepsilon)^{c_{1} m_{n}^{\omega}}\rceil;\\&\quad  s=1, \dots,\lceil (1 / \varepsilon)^{c_{2}s_{n}^{\kappa}}\rceil\bigg\}
	\end{split}
\end{equation*}
 such that for every $m(\xi ; \mathcal{W})-m(\xi_{0}; \mathcal{W}) \in \mathcal{F}_{n}$, there is an index $(\ell, i, s)$ that makes    $m(\xi ; \mathcal{W})-m(\xi_{0n}; \mathcal{W})  \in\left[m_{\ell,i, s}^{L}(\mathcal{W}), m_{\ell,i, s}^{U}(\mathcal{W})\right]$ for any sample point $\mathcal{W}$. Specifically, the brackets  are defined as
 $$
	\begin{aligned}
		m_{\ell,i, s}^{L}(\mathcal{W})=\bigg\{& \Delta g_{s}^{L}\Big(r_{\theta_{0}}-\mu(U, \theta_{\ell,i}-\theta_{0})+c_{\ell, i,s,r_{\theta_{0}}}^{L}  \varepsilon\Big) \\
		&\left.-\int_{a}^{b} 1\left(r_{\theta_{0}} \geq t\right) \exp \left\{g_{s}^{U}\left(t_{\theta_{\ell,i}}+c_{\ell,i,s,t}^{U} \varepsilon\right)\right\} d t\right\} \\
		&-m\left(\xi_{0} ; \mathcal{W}\right),
	\end{aligned}
	$$
	and
	$$
	\begin{aligned}
	m_{\ell,i, s}^{U}(\mathcal{W})=\bigg\{& \Delta g_{s}^{U}\left(r_{\theta_{0}}-\mu(U, \theta_{\ell,i}-\theta_{0})+c_{\ell, i,s,r_{\theta_{0}}}^{U}  \varepsilon\right) \\
	&\left.-\int_{a}^{b} 1\left(r_{\theta_{0}} \geq t\right) \exp \left\{g_{s}^{L}\left(t_{\theta_{\ell,i}}+c_{\ell,i,s,t}^{L} \varepsilon\right)\right\} d t\right\} \\
	&-m\left(\xi_{0} ; \mathcal{W}\right).
	\end{aligned}
	$$
	
  To show that the brackets could introduce a $\varepsilon$-bracket for $\mathcal{F}_n$, we study
	$$
		\left|m_{\ell,i, s}^{U}(\mathcal{W})-m_{\ell,i, s}^{L}(\mathcal{W})\right|	\leq A_1+A_2,
	$$
	where
	$$
	\begin{aligned}
		 &A_1=
		\;\Big|\; g_{s}^{U}\left(r_{\theta_{0}}-\mu(U, \theta_{\ell,i}-\theta_{0})+c_{\ell, i,s,r_{\theta_{0}}}^{U}  \varepsilon\right)
	 - g_{s}^{L}\left(r_{\theta_{0}}-\mu(U, \theta_{\ell,i}-\theta_{0})+c_{\ell, i,s,r_{\theta_{0}}}^{L}  \varepsilon\right)\Big| \\&A_2=\int_{a}^{b} \left| \exp \left\{g_{s}^{U}\left(t_{\theta_{\ell,i}}+c_{\ell,i,s,t}^{U} \varepsilon\right)\right\} -\exp \left\{g_{s}^{L}\left(t_{\theta_{\ell,i}}+c_{\ell,i,s,t}^{L} \varepsilon\right)\right\}
	 \right| dt.
     \end{aligned}
	$$
	For $A_{1}$, according to Taylor expansion, we can obtain
	$$
	\begin{aligned}
		A_{1} \leq &\left|g_{s}^{U}\left(r_{\theta_{0}}-\mu(U, \theta_{\ell,i}-\theta_{0})+c_{\ell, i,s,r_{\theta_{0}}}^{U}  \varepsilon\right) -g\left(r_{\theta_{0}}-\mu(U, \theta_{\ell,i}-\theta_{0})+c_{\ell, i,s,r_{\theta_{0}}}^{U}  \varepsilon\right)\right| \\
		&+\left|g\left(r_{\theta_{0}}-\mu(U, \theta_{\ell,i}-\theta_{0})+c_{\ell, i,s,r_{\theta_{0}}}^{U}  \varepsilon\right)-g\left(r_{\theta_{0}}-\mu(U, \theta_{\ell,i}-\theta_{0})+c_{\ell, i,s,r_{\theta_{0}}}^{L}  \varepsilon\right)\right| \\
		&+\left|g\left(r_{\theta_{0}}-\mu(U, \theta_{\ell,i}-\theta_{0})+c_{\ell, i,s,r_{\theta_{0}}}^{L}  \varepsilon\right)-g_{s}^{L}\left(r_{\theta_{0}}-\mu(U, \theta_{\ell,i}-\theta_{0})+c_{\ell, i,s,r_{\theta_{0}}}^{L}  \varepsilon\right)\right| \\
		\leq &\left\|g_{s}^{U}-g\right\|_{\infty}+ \left\|\dot{g}\right\|_{\infty} \left|c_{\ell, i,s,r_{\theta_{0}}}^{U}- c_{\ell, i,s,r_{\theta_{0}}}^{L}\right|\varepsilon+\left\|g-g_{s}^{L}\right\|_{\infty} \\
		\leq &2\left\|g_{s}^{U}-g_{s}^{L}\right\|_{\infty}+C_{1}\left|c_{\ell, i,s,r_{\theta_{0}}}^{U}- c_{\ell, i,s,r_{\theta_{0}}}^{L}\right|\varepsilon \\
		\leq & 2 \varepsilon+2 C_{1} C\varepsilon \lesssim \varepsilon.
	\end{aligned}
	$$
 The third inequality holds because   $\dot{g}$ is bounded by $C_{1}$, which  may be proportional to $c_{n}$ that is allowed to grow with $n$ slowly enough. For simplicity,  $c_{n}$  is dropped, since   it does not affect the later calculations (see  \cite{Shen1994}, page 591, for their constant $l_{n}$).  For $A_{2}$, by using a similar argument, we have
	$$
	\begin{aligned}
		A_{2} \leq &\int_{a}^{b} \Big\{\; \big| \exp \{g_{s}^{U}(t_{\theta_{\ell,i}}+c_{\ell,i,s,t}^{U} \varepsilon)\} -
		\exp \{g(t_{\theta_{\ell,i}}+c_{\ell,i,s,t}^{U} \varepsilon)\}\big|
		\\&\qquad+\big|\exp \{g(t_{\theta_{\ell,i}}+c_{\ell,i,s,t}^{U} \varepsilon)\}-\exp \{g(t_{\theta_{\ell,i}}+c_{\ell,i,s,t}^{L} \varepsilon)\}  \big|
		\\&\qquad+\big|\exp \{g(t_{\theta_{\ell,i}}+c_{\ell,i,s,t}^{L} \varepsilon)\} -\exp \{g_{s}^{L}(t_{\theta_{\ell,i}}+c_{\ell,i,s,t}^{L} \varepsilon)\}
		\big|\;\Big\} dt \\
		\leq&\int_{a}^{b}\Big|\exp \{g(t_{\theta_{\ell,i}}+c_{\ell,i,s,t}^{U} \varepsilon)\}\|g_{s}^{U}-g\|_{\infty}\exp\{\|g_{s}^{U}-g\|_{\infty}\}
		\\&\qquad+\|\exp\{g\}\dot{g}\|_{\infty}|c_{\ell,i,s,t}^{U} -c_{\ell,i,s,t}^{L} |\varepsilon
         \\&\qquad+\exp \{g(t_{\theta_{\ell,i}}+c_{\ell,i,s,t}^{L} \varepsilon)\}\|g-g_{s}^{L}\|_{\infty}\exp\{\|g-g_{s}^{L}\|_{\infty}\}\Big|dt\\
		\lesssim &\|g_{s}^{U}-g\|_{\infty}+|c_{\ell,i,s,t}^{U} -c_{\ell,i,s,t}^{L} |\varepsilon+\|g-g_{s}^{L}\|_{\infty} \lesssim \varepsilon.
	\end{aligned}
	$$
    It follows that $\|m_{\ell,i, s}^{U}(\mathcal{W})-m_{\ell,i, s}^{L}(\mathcal{W})\|_{\infty} \lesssim \varepsilon$. Therefore, the $\varepsilon$-bracketing number associated with the supremum norm for the class $\mathcal{F}_{n}$ satisfies
	$$
	N_{[\;]}\left(\varepsilon, \mathcal{F}_{n},\|\cdot\|_{\infty}\right) \leq(1 / \varepsilon)^{c_{1} m_{n}^{\omega}}  (1 / \varepsilon)^{c_2s_{n}^{\kappa}}c_3(1/\varepsilon)^{p} \lesssim(1 / \varepsilon)^{c_{1} m_{n}^{\omega}+c_2s_{n}^{\kappa}+p},
	$$
	which completes the proof.
\end{proof}

\subsection{Proof of Lemma \ref{lem4}}\label{pf:lem4}
\begin{proof}
	First, we define  the following  classes of functions
	\begin{equation}\label{space definition}
		\begin{split}
			\mathcal{B}(\eta) &=\left\{\alpha \in \mathcal{B},\left|\alpha-\alpha_{0}\right| \leq \eta\right\} ,
			\\	\mathcal{F}_{n}^{\omega}(\eta)&=\left\{\beta \in \mathcal{F}_{n}^{\omega},\left\|\beta-\beta_{0}\right\|\leq \eta\right\} ,
			\\\mathcal{G}_{n}^{\kappa}(\eta)&=\left\{g \in \mathcal{G}_{n}^{\kappa},\left\|g-g_{0}\right\|_{\mathcal{G}} \leq \eta\right\},
			\\\mathcal{G}_{n}^{\kappa-1}(\eta)&=\left\{  \dot{g} \in \mathcal{G}_{n}^{\kappa-1},\left\|\dot{g} -\dot{g} _{0}\right\|_{\mathcal{G}} \leq \eta\right\}.
		\end{split}
	\end{equation}
Then, it follows by the calculation of \cite{Shen1994} on page 597 that
$$
N_{[\;]}\left(\varepsilon, \mathcal{F}_{n}^{\omega}(\eta),\|\cdot\|_{\infty}\right) \leq(\eta / \varepsilon)^{a_{1} m_{n}^{\omega}} ,\; N_{[\;]}\left(\varepsilon,\mathcal{G}_{n}^{\kappa}(\eta),\|\cdot\|_{\infty}\right) \leq(\eta / \varepsilon)^{a_{2} s_{n}^{\kappa}},
$$
and  $
N_{[\;]}\left(\varepsilon,\mathcal{G}_{n}^{\kappa-1}(\eta),\|\cdot\|_{\infty}\right) \leq(\eta / \varepsilon)^{a_{3} s_{n}^{\kappa}}$ for some constants $a_{1}, a_{2}, a_3>0 .$ From  Condition \ref{assume1} that $\mathcal{B}$ is compact, the covering number for $\mathcal{B}(\eta)$ follows $N\left(\varepsilon, \mathcal{B}(\eta),\|\cdot\|_{\infty}\right) \leq a_{4}(\eta / \varepsilon)^{p}$ for a constant $a_4>0$.
By using a similar argument as   \ref{pf:lem3},   for any $\theta$, there exists $\theta_{\ell,i}=(\alpha_{\ell}, \beta_{i}^{U})$ and constant $C$ such that
	\begin{equation*}
		t_{\theta_{\ell,i}}-C\varepsilon	\leq t_{\theta}\leq  t_{\theta_{\ell,i}}+C\varepsilon
	\end{equation*}
for some $ i\in\{1,\dots,\lceil(\eta/ \varepsilon)^{a_{1} m_{n}^{\omega}}\rceil\}$  and $\ell \in\{1,\dots,  \lceil a_{4}(\eta / \varepsilon)^{p}\rceil\}.$  Let $g_{s}^{L}$ and $g_{s}^{U}$ be the
	functions that bracket $g$ with $\|g_{s}^{U}-g_{s}^{L}\|_{\infty} \leq \varepsilon$; let $\dot{g}_{j}^{L}$ and $\dot{g}_{j}^{U}$ be the  functions that bracket
	$\dot{g}$ with $ \|\dot{g}_{j}^{U}-\dot{g}_{j}^{L} \|_{\infty} \leq \varepsilon$; let
	    $g_{s}^{L}(t_{\theta_{\ell,i}}+a_{\ell,i,s,t}^{1,L} \varepsilon)$ and $g_{s}^{U}(t_{\theta_{\ell,i}}+a_{\ell,i,s,t}^{1,U} \varepsilon)$ be the minimum and maximum
	values of $g_{s}^{L}$ and $g_{s}^{U}$ within the closed interval $[t_{\theta_{\ell,i}}-C \varepsilon, t_{\theta_{\ell,i}}+C \varepsilon]$, where  $a_{\ell,i,s,t}^{1,L}$ and $a_{\ell,i,s,t}^{1,U}$ are
	two constants that have its absolute value bounded by $C$; let    $\dot{g}_{j}^{L}(t_{\theta_{\ell,i}}+a_{\ell,i,j,t}^{2,L} \varepsilon)$ and $\dot{g}_{j}^{U}(t_{\theta_{\ell,i}}+a_{\ell,i,j,t}^{2,U} \varepsilon)$ be the minimum and maximum
	values of $\dot{g}_{j}^{L}$ and $\dot{g}_{j}^{U}$ within   $[t_{\theta_{\ell,i}}-C \varepsilon, t_{\theta_{\ell,i}}+C \varepsilon]$, where $a_{\ell,i,j,t}^{2,L}$ and $a_{\ell,i,j,t}^{2,U}$ are
	two constants that have its absolute value bounded by $C$; and let    $h_3(t_{\theta_{\ell,i}}+a_{\ell,i,t}^{3,L}\varepsilon)$ and $h_3(t_{\theta_{\ell,i}}+a_{\ell,i,t}^{3,U}\varepsilon)$ be  the minimum and maximum values
	of $h_{3}$ within $[t_{\theta_{\ell,i}}-C \varepsilon, t_{\theta_{\ell,i}}+C \varepsilon],$ where $a_{\ell,i,t}^{3,L}$ and $a_{\ell,i,t}^{3,U}$ are two
	constants that have its absolute value bounded by $C$.  Then, we can   construct a set of brackets
	\begin{equation}\label{brae}
		\begin{split}
			\bigg\{\left[\phi_{\ell,i,j,s}^{L}(\mathcal{W};\bh),\phi_{\ell,i,j,s}^{U}(\mathcal{W};\bh)\right]&: \ell=1,\dots \lceil a_{4}(\eta / \varepsilon)^{p}\rceil;  i=1, \dots,\lceil(\eta / \varepsilon)^{a_{1} m_{n}^{\omega}}\rceil;\\&\quad
			j=1, \dots,\lceil (\eta / \varepsilon)^{a_{3}s_{n}^{\kappa}}\rceil;
			 s=1, \dots,\lceil (\eta/ \varepsilon)^{a_{2}s_{n}^{\kappa}}\rceil\bigg\}
		\end{split}
	\end{equation}
    such that for every element $\dot{m}(\xi;\mathcal{W})[\bh]-\dot{m}(\xi_{0};\mathcal{W})[\bh]\in\mathcal{F}_{n}(\bh,\eta)$,  there is an index $(\ell,i,j,s)$ that makes
	$$
\dot{m}(\xi;\mathcal{W})[\bh]-\dot{m}(\xi_{0};\mathcal{W})[\bh] \in\left[\phi_{\ell,i,j,s}^{L}(\mathcal{W}; \bh),\phi_{\ell,i,j,s}^{U}(\mathcal{W};\bh)\right]
	$$
	for any sample point $\mathcal{W}.$  To introduce the definitions of the brackets, we first define that   for any function $f(t)$,  $f^{+}(t)=\max\{f(t),0\}$ and $f^{-}(t)=\max\{-f(t),0
\}$, so that both $f^{+}$ and $f^{-}$ are nonnegative and   $f^{+}-f^{-}=f.$  Let
\begin{equation*}
	\begin{split}
		 \zeta_{\ell,i,j,s}^{L}&=\int_{a}^{b}I(r_{\theta_{0}}\geq t )  \exp\{g_{s}^{L}(t_{\theta_{\ell,i}}+a_{\ell,i,s,t}^{1,L} \varepsilon)\}\{\dot{g}_{j}^{L}\}^{+}(t_{\theta_{\ell,i}}+a_{\ell,i,j,t}^{2,L} \varepsilon)
		dt
		\\&\quad-\int_{a}^{b}I(r_{\theta_{0}}\geq t )  \exp\{g_{s}^{U}(t_{\theta_{\ell,i}}+a_{\ell,i,s,t}^{1,U} \varepsilon)\}\{\dot{g}_{j}^{L}\}^{-}(t_{\theta_{\ell,i}}+a_{\ell,i,j,t}^{2,L} \varepsilon)
		dt,
		\\	 \zeta_{\ell,i,j,s}^{U}&=\int_{a}^{b}I(r_{\theta_{0}}\geq t )  \exp\{g_{s}^{U}(t_{\theta_{\ell,i}}+a_{\ell,i,s,t}^{1,U} \varepsilon)\}\{\dot{g}_{j}^{U}\}^{+}(t_{\theta_{\ell,i}}+a_{\ell,i,j,t}^{2,U} \varepsilon)
		dt
		\\&\quad-\int_{a}^{b}I(r_{\theta_{0}}\geq t )  \exp\{g_{s}^{L}(t_{\theta_{\ell,i}}+a_{\ell,i,s,t}^{1,L} \varepsilon)\}\{\dot{g}_{j}^{U}\}^{-}(t_{\theta_{\ell,i}}+a_{\ell,i,j,t}^{2,U} \varepsilon)
		dt.
	\end{split}
\end{equation*}
Then, it follows that
\begin{equation*}
 \zeta_{\ell,i,j,s}^{L}	\leq \int_{a}^{b}I(r_{\theta_{0}}\geq t )  \exp\{g(t_{\theta})\}\dot{g}(t_{\theta})
	dt\leq \zeta_{\ell,i,j,s}^{U},
\end{equation*}
which further leads to
\begin{equation*}
	\begin{split}
			\gamma_{\ell,i,j,s}^{L}	&\equiv  \zeta_{\ell,i,j,s}^{L}-\Delta\dot{g}_{j}^{U}(r_{\theta_{0}}-\mu(U,\theta_{\ell,i}-\theta_{0})+ a_{\ell,i,j,r_{\theta_{0}}}^{2,U} \varepsilon)\\&\leq \int_{a}^{b}I(r_{\theta_{0}}\geq t )  \exp\{g(t_{\theta})\}\dot{g}(t_{\theta})
			dt-\Delta\dot{g}(r_{\theta})
			\\&\leq \zeta_{\ell,i,j,s}^{U} - \Delta\dot{g}_{j}^{L}(r_{\theta_{0}}-\mu(U,\theta_{\ell,i}-\theta_{0})+ a_{\ell,i,j,r_{\theta_{0}}}^{2,L} \varepsilon)\equiv	\gamma_{\ell,i,j,s}^{U}	
	\end{split}
\end{equation*}
Let $\delta=(h_1, h_2)$ and $r_{\theta_{\ell,i}}=r_{\theta_{0}}-\mu(U,\theta_{\ell,i} -\theta_{0})$. For each $(\ell,i,j,s),$ the  bracket in \eqref{brae} is defined as
	$$
	\begin{aligned}
	\phi_{\ell,i,j,s}^{L}(\mathcal{W}; \bh)
	 &=\gamma_{\ell,i,j,s}^{L}\cdot\mu(U,\delta)^{+}
	 -\gamma_{\ell,i,j,s}^{U}\cdot\mu(U,\delta)^{-}
+\Delta h_3\big(r_{\theta_{\ell,i}}+a_{\ell,i,r_{\theta_{0}}}^{3,L}\varepsilon\big) \\&\quad-\int_{a}^{b}I(r_{\theta_{0}}\geq t)\{h_3\}^{+}(t_{\theta_{\ell,i}}+a_{\ell,i,t}^{3,U}\varepsilon)\exp \{g_{s}^{U}(t_{\theta_{\ell,i}}+a_{\ell,i,s,t}^{1,U} \varepsilon)\}dt
 \\&\quad+\int_{a}^{b}I(r_{\theta_{0}}\geq t)\{h_3\}^{-}(t_{\theta_{\ell,i}}+a_{\ell,i,t}^{3,U}\varepsilon)\exp \{g_{s}^{L}(t_{\theta_{\ell,i}}+a_{\ell,i,s,t}^{1,L} \varepsilon)\}dt
 	\\&\quad -\dot{m}(\xi_{0};\mathcal{W})[\bh]
 	\end{aligned}
	$$
	and
		$$
	\begin{aligned}
		\phi_{\ell,i,j,s}^{U}(\mathcal{W};\bh)
		&=\gamma_{\ell,i,j,s}^{U}\cdot\mu(U,\delta)^{+}
		-\gamma_{\ell,i,j,s}^{L}\cdot\mu(U,\delta)^{-}
		+\Delta h_3\big(r_{\theta_{\ell,i}}+a_{\ell,i,r_{\theta_{0}}}^{3,U}\varepsilon\big) \\&\quad-\int_{a}^{b}I(r_{\theta_{0}}\geq t)\{h_3\}^{+}(t_{\theta_{\ell,i}}+a_{\ell,i,t}^{3,L}\varepsilon)\exp \{g_{s}^{L}(t_{\theta_{\ell,i}}+a_{\ell,i,s,t}^{1,L} \varepsilon)\}dt
		\\&\quad+\int_{a}^{b}I(r_{\theta_{0}}\geq t)\{h_3\}^{-}(t_{\theta_{\ell,i}}+a_{\ell,i,t}^{3,L}\varepsilon)\exp \{g_{s}^{U}(t_{\theta_{\ell,i}}+a_{\ell,i,s,t}^{1,U} \varepsilon)\}dt
			\\&\quad -\dot{m}(\xi_{0};\mathcal{W})[\bh].
	\end{aligned}
	$$
	Following a similar argument in \ref{pf:lem3}, we can show that
	$ \|\phi_{\ell,i,j,s}^{U}(\mathcal{W}; \bh)-\phi_{\ell,i,j,s}^{L}(\mathcal{W}; \bh) \|_{\infty}\lesssim \varepsilon.$ Therefore, the $\varepsilon$-bracketing number associated with the supremum norm for the class $\mathcal{F}_{n}(\bh,\eta)$ satisfies
	\begin{equation*}
	 \mathcal{N}_{[\;]}(\varepsilon, \mathcal{F}_{n}(\bh,\eta),\|\cdot\|_{\infty})\lesssim  (\eta/\varepsilon)^{c_3m_{n}^{\omega}+c_4s_{n}^{\kappa}+p}
	\end{equation*}
for some constants $c_3,c_4>0$, which completes our proof.
\end{proof}

\subsection{Proof of Lemma \ref{lem5}}
\begin{proof}
 Let  spaces $\mathcal{B}(\eta), \mathcal{F}_{n}^{\omega}(\eta), \mathcal{G}_{n}^{\kappa}(\eta)$ and $ \mathcal{G}_{n}^{\kappa-1}(\eta)$ have the same definitions as \eqref{space definition} in \ref{pf:lem4}.  Define
	\begin{equation*}
		\begin{split}
			&\mathcal{F}_{n}^{\omega}(h_2,\eta)=\big\{\tilde{h}_2 \in\mathcal{F}_{n}^{\omega},\|\tilde{h}_2-h_2\|_{\infty}\leq \eta\big\} ,
			\\&\mathcal{G}_{n}^{\kappa}(h_3,\eta)=\big\{\tilde{h}_3 \in \mathcal{G}_{n}^{\kappa},\|\tilde{h}_3-h_3\|_{\infty} \leq \eta\big\}.
		\end{split}
	\end{equation*}
	Then,  it follows by the calculation of \cite{Shen1994} on page 597 that
	$$
	N_{[\;]}\left(\varepsilon,\mathcal{F}_{n}^{\omega}(h_2,\eta),\|\cdot\|_{\infty}\right) \leq(\eta / \varepsilon)^{a_{5} m_{n}^{\omega}},\; N_{[\;]}\left(\varepsilon,\mathcal{G}_{n}^{\kappa}(h_3,\eta),\|\cdot\|_{\infty}\right) \leq(\eta / \varepsilon)^{a_{6} s_{n}^{\kappa}}.
	$$
	for some constants $a_{5}, a_{6}>0.$   Following the same argument as that in \ref{pf:lem4},  for any $\xi=(\alpha,\beta, g)$, there exists $\theta_{\ell,i}=(\alpha_{\ell}, \beta_{i}^{U})$ and constant $C$ such that $ t_{\theta_{\ell,i}}-C\varepsilon	\leq t_{\theta}\leq  t_{\theta_{\ell,i}}+C\varepsilon,
$
	\begin{equation*}
		\begin{split}
			\gamma_{\ell,i,j,s}^{L}	 \leq \int_{a}^{b}I(r_{\theta_{0}}\geq t )  \exp\{g(t_{\theta})\}\dot{g}(t_{\theta})
			dt-\Delta\dot{g}(r_{\theta})
			 \leq  \gamma_{\ell,i,j,s}^{U},
		\end{split}
	\end{equation*}
and
\begin{equation*}
g_{s}^{L}(t_{\theta_{\ell,i}}+a_{\ell,i,s,t}^{1,L} \varepsilon)\leq g(t_{\theta})\leq g_{s}^{U}(t_{\theta_{\ell,i}}+a_{\ell,i,s,t}^{1,U} \varepsilon)
\end{equation*}
	for  some  $\ell\in\{1,\dots \lceil a_{4}(\eta / \varepsilon)^{p}\rceil\}, i\in\{1, \dots,\lceil(\eta / \varepsilon)^{a_{1} m_{n}^{\omega}}\rceil\},	j\in\{1, \dots,\lceil (\eta / \varepsilon)^{a_{3}s_{n}^{\kappa}}\rceil\}$ and $s\in\{1, \dots,\lceil (\eta/ \varepsilon)^{a_{2}s_{n}^{\kappa}}\rceil\}.$ Let $\tilde{h}_{2k}^{L}$ and $\tilde{h}_{2k}^{U}$ be the  functions that bracket
	$\tilde{h}_2$ with $ \|\tilde{h}_{2k}^{U}-\tilde{h}_{2k}^{L}\|_{\infty} \leq \varepsilon$;  let $\tilde{h}_{3d}^{L}$ and $\tilde{h}_{3d}^{U}$ be the  functions that bracket
	$\tilde{h}_3$ with $ \|\tilde{h}_{3d}^{U}-\tilde{h}_{3d}^{L}\|_{\infty} \leq \varepsilon$;  and let
	$\tilde{h}_{3d}^{L}(t_{\theta_{\ell,i}}+a_{\ell,i,d,t}^{4,L} \varepsilon)$ and $\tilde{h}_{3d}^{U}(t_{\theta_{\ell,i}}+a_{\ell,i,d,t}^{4,U} \varepsilon)$ be the minimum and maximum
	values of $\tilde{h}_{3d}^{L}$ and $\tilde{h}_{3d}^{U}$ within the closed interval $[t_{\theta_{\ell,i}}-C \varepsilon, t_{\theta_{\ell,i}}+C \varepsilon]$, where  $a_{\ell,i,d,t}^{4,L}$ and $a_{\ell,i,d,t}^{4,U}$ are
	two constants that have its absolute values bounded by $C$. Based on Condition \ref{assume2}, we have
	\begin{equation*}
 \int_{0}^{1}\Big| \big\{h_2(s)-\tilde{h}_{2}(s)\big\}Z(s)-\big\{h_2(s)-\tilde{h}_{2k}^{U}(s)\big\}Z(s)\Big| ds\leq c_7\varepsilon
	\end{equation*}
	for a constant $ c_7>0$. Let  $\delta_{k}=(0,h_2-\tilde{h}_{2k}^{U})$. It follows that
	\begin{equation*}
		\begin{split}
		\eta_{k}^{L}\equiv\mu(U, \delta_{k})-c_7\varepsilon\leq \int_{0}^{1} \big \{h_2(s)-\tilde{h}_{2}(s)\big\}Z(s)ds\leq
			\mu(U, \delta_{k})+c_7\varepsilon\equiv\eta_{k}^{U}.
		\end{split}
	\end{equation*}
	We can construct a set of brackets
	\begin{equation*}
		\begin{split}
			\bigg\{\left[f_{\ell,i,j,s,k,d}^{L}(\mathcal{W};\bh),f_{\ell,i,j,s,k,d}^{U}(\mathcal{W};\bh)\right]:& \ell=1,\dots \lceil a_{4}(\eta / \varepsilon)^{p}\rceil;  i=1, \dots,\lceil(\eta / \varepsilon)^{a_{1} m_{n}^{\omega}}\rceil;\\&
			j=1, \dots,\lceil (\eta / \varepsilon)^{a_{3}s_{n}^{\kappa}}\rceil;
			s=1, \dots,\lceil (\eta/ \varepsilon)^{a_{2}s_{n}^{\kappa}}\rceil;
			\\&
			k=1,\dots\lceil (\eta / \varepsilon)^{a_{5}m_{n}^{\omega}}\rceil;
			d=1,\dots\lceil (\eta / \varepsilon)^{a_{6}s_{n}^{\kappa}}\rceil
			\bigg\}
		\end{split}
	\end{equation*}
	such that for every element $\dot{m}(\xi;\mathcal{W})[\bh-\tilde{\bh}] \in\mathcal{L}_{n}(\textbf{h},\eta)$, there is an index $(\ell,i,j,s,k,d)$   that makes
	$$
	\dot{m}(\xi;\mathcal{W})[\bh-\tilde{\bh}]  \in\left[f_{\ell,i,j,s,k,d}^{L}(\mathcal{W};\bh),f_{\ell,i,j,s,k,d}^{U}(\mathcal{W};\bh)\right]
	$$
	for any sample point $\mathcal{W}$.
Specifically, for each $(\ell,i,j,s,k,d),$ the bracket is defined as
	$$
\begin{aligned}
f_{\ell,i,j,s,k,d}^{L}(\mathcal{W};\bh)
	&=\{\gamma_{\ell,i,j,s}^{L}\}^{+}\{\eta_{k}^{L}\}^{+}-\{\gamma_{\ell,i,j,s}^{L}\}^{-}\{\eta_{k}^{U}\}^{+}
-\{\gamma_{\ell,i,j,s}^{U}\}^{+}\{\eta_{k}^{L}\}^{-}+\{\gamma_{\ell,i,j,s}^{U}\}^{-}\{\eta_{k}^{U}\}^{-}
	\\&	\quad +\Delta h_3\big(r_{\theta_{\ell,i}}+a_{\ell,i,r_{\theta_{0}}}^{3,L}\varepsilon\big)-\Delta \tilde{h}_{3d}^{U}\big(r_{\theta_{\ell,i}}+a_{\ell,i,d,r_{\theta_{0}}}^{4,U}\varepsilon\big) \\&\quad-\int_{a}^{b}I(r_{\theta_{0}}\geq t)\{h_3\}^{+}(t_{\theta_{\ell,i}}+a_{\ell,i,t}^{3,U}\varepsilon)\exp \{g_{s}^{U}(t_{\theta_{\ell,i}}+a_{\ell,i,s,t}^{1,U} \varepsilon)\}dt
	\\&\quad+\int_{a}^{b}I(r_{\theta_{0}}\geq t)\{h_3\}^{-}(t_{\theta_{\ell,i}}+a_{\ell,i,t}^{3,U}\varepsilon)\exp \{g_{s}^{L}(t_{\theta_{\ell,i}}+a_{\ell,i,s,t}^{1,L} \varepsilon)\}dt
\\&\quad-\int_{a}^{b}I(r_{\theta_{0}}\geq t)\{\tilde{h}_{3d}^{L}\}^{-}(t_{\theta_{\ell,i}}+a_{\ell,i,d,t}^{4,L}\varepsilon)\exp \{g_{s}^{U}(t_{\theta_{\ell,i}}+a_{\ell,i,s,t}^{1,U} \varepsilon)\}dt
\\&\quad+\int_{a}^{b}I(r_{\theta_{0}}\geq t)\{\tilde{h}_{3d}^{L}\}^{+}(t_{\theta_{\ell,i}}+a_{\ell,i,d,t}^{4,L}\varepsilon)\exp \{g_{s}^{L}(t_{\theta_{\ell,i}}+a_{\ell,i,s,t}^{1,L} \varepsilon)\}dt.
\end{aligned}
$$
and
	$$
\begin{aligned}
	f_{\ell,i,j,s,k,d}^{U}(\mathcal{W};\bh)
	&=\{\gamma_{\ell,i,j,s}^{U}\}^{+}\{\eta_{k}^{U}\}^{+}-\{\gamma_{\ell,i,j,s}^{U}\}^{-}\{\eta_{k}^{L}\}^{+}
	-\{\gamma_{\ell,i,j,s}^{L}\}^{+}\{\eta_{k}^{U}\}^{-}+\{\gamma_{\ell,i,j,s}^{L}\}^{-}\{\eta_{k}^{L}\}^{-}
	\\&	\quad +\Delta h_3\big(r_{\theta_{\ell,i}}+a_{\ell,i,r_{\theta_{0}}}^{3,U}\varepsilon\big)-\Delta \tilde{h}_{3d}^{L}\big(r_{\theta_{\ell,i}}+a_{\ell,i,d,r_{\theta_{0}}}^{4,L}\varepsilon\big)
	\\&\quad-\int_{a}^{b}I(r_{\theta_{0}}\geq t)\{h_3\}^{+}(t_{\theta_{\ell,i}}+a_{\ell,i,t}^{3,L}\varepsilon)\exp \{g_{s}^{L}(t_{\theta_{\ell,i}}+a_{\ell,i,s,t}^{1,L} \varepsilon)\}dt
	\\&\quad+\int_{a}^{b}I(r_{\theta_{0}}\geq t)\{h_3\}^{-}(t_{\theta_{\ell,i}}+a_{\ell,i,t}^{3,L}\varepsilon)\exp \{g_{s}^{U}(t_{\theta_{\ell,i}}+a_{\ell,i,s,t}^{1,U} \varepsilon)\}dt
	\\&\quad-\int_{a}^{b}I(r_{\theta_{0}}\geq t)\{\tilde{h}_{3d}^{U}\}^{-}(t_{\theta_{\ell,i}}+a_{\ell,i,d,t}^{4,U}\varepsilon)\exp \{g_{s}^{L}(t_{\theta_{\ell,i}}+a_{\ell,i,s,t}^{1,L} \varepsilon)\}dt
	\\&\quad+\int_{a}^{b}I(r_{\theta_{0}}\geq t)\{\tilde{h}_{3d}^{U}\}^{+}(t_{\theta_{\ell,i}}+a_{\ell,i,d,t}^{4,U}\varepsilon)\exp \{g_{s}^{U}(t_{\theta_{\ell,i}}+a_{\ell,i,s,t}^{1,U} \varepsilon)\}dt
\end{aligned}
$$
	Following a similar argument in \ref{pf:lem3}, we can show that
	$ \|f_{\ell,i,j,s,k,d}^{L}(\mathcal{W};\bh)-f_{\ell,i,j,s,k,d}^{U}(\mathcal{W};\bh)\|_{\infty}\lesssim \varepsilon.$ Therefore, the $\varepsilon$-bracketing number associated with the supremum norm for the class $\mathcal{L}_{n}(\textbf{h},\eta)$ satisfies
\begin{equation*}
	\mathcal{N}_{[\;]}(\varepsilon, \mathcal{L}_{n}(\textbf{h},\eta),\|\cdot\|_{\infty})\lesssim  (\eta/\varepsilon)^{c_5m_{n}^{\omega}+c_6s_{n}^{\kappa}+p}
\end{equation*}
for some constants $c_5,c_6>0,$ which completes our proof.
\end{proof}

\newpage

\begin{table}[H]
	\centering
	\caption{Simulation results for the proposed estimate of $\theta$. (a) Exponential; (b) Gaussian Mixture; (c) Extreme Value.}
	\label{table_1}
	\begin{tabular}{cccccccccccc} \hline
		Err.               &      &      &\multicolumn{4}{c}{Censoring Rate $=25\%$} & &\multicolumn{4}{c}{Censoring Rate $=40\%$}\\  \cline{4-7}\cline{9-12}
		dist                &  $n$ &                & BIAS  & SSE   & ESE    & CP     & & BIAS  & SSE   & ESE    & CP    \\ \cline{1-12}
		(a)                  &  400 & $\alpha _1$    &-0.0033&0.1189 & 0.1162 & 0.942 & & -0.0089& 0.1385& 0.1307& 0.938 \\
		&      & $\alpha _2$    &-0.0017&0.1168 & 0.1172 & 0.952 & & -0.0071& 0.1377& 0.1316& 0.937 \\
		&  600 & $\alpha _1$    &-0.0053&0.0969 & 0.0944 & 0.941 & & -0.0113& 0.1112& 0.1062& 0.942 \\
		&      & $\alpha _2$    &0.0006 &0.0983 & 0.0950 & 0.938 & & -0.0035& 0.1124& 0.1066& 0.941 \\
		&  800 & $\alpha _1$    &-0.0020&0.0813 & 0.0814 & 0.941 & & -0.0002& 0.0984& 0.0949& 0.945 \\
		&      & $\alpha _2$    &-0.0012&0.0815 & 0.0819 & 0.943 & & -0.0015& 0.0959& 0.0952& 0.946 \\ \hline
		(b)                  &  400 & $\alpha _1$    &0.0092 &0.2287 & 0.2187 & 0.939 & & -0.0158& 0.2385& 0.2304& 0.943 \\
		&      & $\alpha _2$    &-0.0053&0.2117 & 0.2197 & 0.960 & & -0.0168& 0.2369& 0.2323& 0.938 \\
		&  600 & $\alpha _1$    &0.0035 &0.1778 & 0.1792 & 0.948 & & -0.0279& 0.1993& 0.1879& 0.934 \\
		&      & $\alpha _2$    &0.0146 &0.1866 & 0.1801 & 0.943 & & -0.0286& 0.1963& 0.189& 0.938 \\
		&  800 & $\alpha _1$    &0.0162 &0.1475 & 0.1497 & 0.956 & & 0.0179& 0.1623& 0.1613& 0.955 \\
		&      & $\alpha _2$    &0.0130 &0.1490 & 0.1503 & 0.955 & & 0.0124& 0.1600& 0.1622& 0.954 \\ \hline
		(c)                  &  400 & $\alpha _1$    &-0.0005&0.1246 & 0.1217 & 0.945 & & $<$0.0001& 0.1280& 0.1256& 0.935 \\
		&      & $\alpha _2$    &-0.0026&0.1191 & 0.1225 & 0.960 & & -0.0002& 0.1225& 0.1264& 0.958 \\
		&  600 & $\alpha _1$    &0.0010 &0.1001 & 0.0993 & 0.945 & & 0.0031& 0.1034& 0.1025& 0.953 \\
		&      & $\alpha _2$    &-0.0050&0.1031 & 0.0997 & 0.948 & & -0.0049& 0.1053& 0.1029& 0.948\\
		&  800 & $\alpha _1$    &-0.0007&0.0832 & 0.0840 & 0.951 & & -0.0002& 0.0863& 0.0877& 0.953 \\
		&      & $\alpha _2$    &-0.0030&0.0845 & 0.0844 & 0.948 & & -0.0010& 0.0881& 0.0883& 0.950 \\ \hline
	\end{tabular}
\end{table}

\begin{table}[H]
	\centering
	\caption{MSE of the proposed estimates of $\beta(\cdot)$ and $g(\cdot)$. (a) Exponential; (b) Gaussian Mixture; (c) Extreme Value.}
	\label{theta_bias}
	\begin{tabular}{ccccccccc} \hline
		Err.                 &      & \multicolumn{3}{c}{Censoring Rate $=25\%$}      & &  \multicolumn{3}{c}{Censoring Rate $=40\%$}     \\ \cline{3-5}\cline{7-9}
		dist                 &  $n$ &  &  $\beta(\cdot)$  & $g(\cdot)$  & & &   $\beta(\cdot)$  & $g(\cdot)$ \\ \cline{1-9}
		(a)                  &  400 &  & 0.0183&0.0198 & &  &  	0.0230 &0.0599 \\
		&  600 &  & 0.0153&0.0138 & &  &	0.0184 &0.0464 \\
		&  800 &  & 0.0118&0.0122 & &  &	0.0145 &0.0233 \\\hline
		(b)                  &  400 &  & 0.0745&0.1645 & &  &   0.0538 &0.1716 \\
		&  600 &  & 0.0580&0.1600 & &  &	0.0409 &0.1710 \\
		&  800 &  & 0.0333&0.0760 & &  &	0.0403 &0.1080 \\\hline
		(c)                  &  400 &  & 0.0220&0.1259 & &  &   0.0218 &0.1279 \\
		&  600 &  & 0.0164&0.1157 & &  &	0.0169 &0.1125 \\
		&  800 &  & 0.0158&0.0980 & &  &	0.0167 &0.0975 \\\hline		
	\end{tabular}
\end{table}

\begin{table}[H]
	\centering
	\caption{Estimation results of regression coefficients of scalar covariates for the SOFA data analysis.\label{sofa_theta}}
	\begin{tabular}{ccccc} \hline
		&  $\hat{\theta}$    & $S.E.$   & $t$-value & $p$-value\\ \cline{1-5}
		Gender (male=1)     & $0.26142$  & $0.04383$ &	$5.96$ & $<0.0001$\\
		Age                 & $-0.02456$  & $0.00016$ &	$-156.66$ & $<0.0001$\\ \hline
	\end{tabular}
\end{table}

\begin{thebibliography}{99}

\bibitem[Buckley and James (1979)]{buck1979}
Buckley, J. and James, I. (1979). Linear regression with censored data. {\it Biometrika} {\bf 66} 429--436.

\bibitem[Cai and Yuan (2012)]{Cai2012}
 Cai, T. T. and Yuan, M. (2012). Minimax and adaptive prediction for functional linear regression. {\it Journal of the American Statistical Association} {\bf 107} 1201--1216.

\bibitem[Cardot, Ferraty, and Sarda (1999)]{Cardot1999}
Cardot, H.,  Ferraty, F.  and Sarda, P. (1999). Functional linear model.  {\it Statistics \& Probability Letters}  {\bf 45} 11--22.

\bibitem[Chen et al. (2011)]{Chen2011}		
Chen, K., Chen, K., M\"{u}ller, H. G. and Wand, J. L. (2011). Stringing high-dimensional data for functional analysis.  {\it Journal of the American Statistical Association}. {\bf 106} 275--284.

\bibitem[Chen (2007)]{Chen2007}
Chen, X. (2007).   Large sample sieve estimation of semi-nonparametric models.   {\it Handbook of econometrics}  {\bf 6} 5549--5632.

\bibitem[Cox (1972)]{Cox1972}
Cox, D. R. (1972). Regression models and life-tables. {\it Journal of the Royal Statistical Society. Series B.	Methodological} {\bf 34} 187--220.

\bibitem[Crainiceanu, Staicu, and Din (2009)]{Crainiceanu2009}
Crainiceanu, C. M.,  Staicu, A. M. and Di, C. Z. (2009). Generalized multilevel functional regression. {\it Journal of the American Statistical Association}  {\bf 104} 1550--1561.

\bibitem[Cui, Crainiceanu, and Leroux (2021)]{Cui2021}
Cui, E., Crainiceanu, C. M. and Leroux, A. (2021). Additive Functional Cox Model. {\it Journal of Computational and Graphical Statistics}  {\bf 0} 1--14.

\bibitem[Ding and Nan (2011)]{Nan2011}
Ding, Y. and Nan, B. (2011).  A sieve {$M$}-theorem for bundled parameters in semiparametric models, with application to the efficient estimation in a linear model for censored data.  {\it The Annals of Statistics}  {\bf 39} 3032--3061.

\bibitem[Elias et al. (2020)]{Elias}
Elias, A., Agbarieh, R., Saliba, W., Khoury, J., Bahouth, F., Nashashibi, J. and Azzam, Z. (2020).
SOFA score and short-term mortality in acute decompensated heart failure.
{\it Scientific Reports} {\bf 10}, 20802.

 \bibitem[Ferraty and Vieu (2006)]{Ferraty2006}
 Ferraty, F. and Vieu, P. (2006). {\it Nonparametric Functional Data Analysis: Theory and Practice}.  Springer, New York.

\bibitem[{Ferreira et al.(2001)}]{Ferreira}
Ferreira, F. L., Bota, D. P., Bross, A., M\'elot, C. and Vincent, J. L. (2001).
Serial evaluation of the SOFA score to predict outcome in critically ill patients. {\it Journal of the American Medical Association} {\bf 286}, 1754–1758.

\bibitem[{Goldsmith et al. (2021)}]{Goldsmith}
Goldsmith, J., Scheipl, F., Huang, L., Wrobel, J., Di, C., Gellar, J., Harezlak, J., McLean M. W., Swihart, B., Xiao, L., Crainiceanu, C. and Reiss, P. T. (2021).
{\it refund: Regression with Functional Data}. R package version 0.1-24.

\bibitem[Hao et al. (2021)]{Hao2020}
Hao, M., Liu, K. Y., Xu, W.  and  Zhao, X. (2021). Semiparametric Inference for the Functional Cox Model. {\it Journal of the American Statistical Association}  {\bf 116} 1319--1329.

\bibitem[Huang (1999)]{Huang1999}
Huang, J. (1999). Efficient estimation of the partly linear additive {C}ox 	model. {\it The Annals of Statistics}  {\bf 27} 1536--1563.

\bibitem[James and  Silverman (2005)]{James2005}
James, G. M. and Silverman, B. W. (2005). Functional adaptive model estimation. {\it Journal of the American Statistical Association}   {\bf 100}  565--576.

\bibitem[Jiang et al. (2020)]{Jiang2020}
Jiang, F., Cheng, Q., Yin, G. and Shen, H. (2020). Functional censored quantile regression.
{\it Journal of the American Statistical Association } {\bf 115} 931--944.

\bibitem[Jin et al. (2003)]{Jin2003}
Jin, Z., Lin, D. Y., Wei, L. J. and Ying, Z. (2003). Rank-based inference for the accelerated
failure time model. {\it Biometrika} {\bf 90}  341--353.

\bibitem[Jin, Lin, and Ying (2006)]{Jin2006}
Jin, Z., Lin, D. Y. and Ying, Z. (2006). On least-squares regression with censored data. {\it Biometrika} {\bf 93}  147--161.


\bibitem[Kong et al.  (2018)]{Kong2018}
Kong, D., Ibrahim, J. G., Lee, E. and  Zhu, H. (2018). F{LCRM}: functional linear {C}ox regression model. {\it Biometrics}  {\bf 74} 109--117.

\bibitem[Kong et al. (2018)]{Nan2018}
Kong, S., Nan, B., Kalbfleisch, J. D.,	Saran, R. and Hirth, R. (2018). Conditional modeling of longitudinal data with terminal event. {\it Journal of the American Statistical Association} {\bf 113} 357--368.


\bibitem[Kuchibhotla and  Patra (2020)]{Kuchibhotla2020}
Kuchibhotla, A. K. and Patra, R. K. (2020). Efficient estimation in single index models through smoothing
splines. {\it Bernoulli}  {\bf 26} 1587--1618.

\bibitem[Lai and Ying (1991a)]{Lai1991a}
Lai, T. L. and Ying, Z. (1991a). Rank regression methods for left-truncated and right-censored data. {\it The Annals of Statistics}
{\bf 19} 531--556.

\bibitem[Lai and Ying (1991b)]{Lai1991b}
Lai, T. L. and Ying, Z. (1991b). Large sample theory of a modified {B}uckley-{J}ames estimator for regression analysis with censored data. {\it The Annals of Statistics} {\bf 19} 1370--1402.

\bibitem[Li, Wang, and  Carroll (2010)]{Li2010}
Li, Y.,  Wang, N.  and Carroll, R. J. (2010). Generalized functional linear models with semiparametric single-index interactions. {\it Journal of the American Statistical Association}  {\bf 105} 621--633.

\bibitem[Lin and Chen (2013)]{Lin2013}
Lin, Y. and Chen, K. (2013). Efficient estimation of the censored linear regression model. {\it Biometrika} {\bf 100}  525-530.



\bibitem[Marx and Eilers (1999)]{Marx1999}
Marx, B. D. and  Eilers, P. H. (1999). Generalized linear regression on sampled signals and curves: a P-spline approach. {\it Technometrics}  {\bf 41} 1--13.

\bibitem[McLean et al. (2014)]{McLean2014}
McLean, M. W., Hooker, G., Staicu, A. M., Scheipl, F. and Ruppert, D. (2014). Functional generalized additive models. {\it Journal of Computational and Graphical Statistics}  {\bf 23} 249--269.

\bibitem[Miller and Halpern (1982)]{miller1982}
Miller, R. and Halpern, J. (1982). Regression with censored data. {\it Biometrika}  {\bf 69} 521--531.

\bibitem[Morris (2015)]{Morris2015}
Morris, J. S. (2015). Functional regression. {\it Annual Review of Statistics and Its Application} {\bf 2}, 321--359.

\bibitem[M\"{u}ller and Stadtm\"{u}ller (2005)]{M2005}
 M\"{u}ller, H. G. and Stadtm\"{u}ller, U. (2005).  Generalized functional linear models. {\it The Annals of Statistics} {\bf 33}, 774--805.

\bibitem[Murphy, van der Vaart, and Wellner (1999)]{Wellner1999}
Murphy, S. A., van der Vaart, A. W. and Wellner, J. A. (1999).  Current status regression.  {\it Mathematical Methods of Statistics} {\bf 8},    407--425.

\bibitem[Needham et al. (2006)]{Need2006}
Needham, D. M., Dennison, C. R., Dowdy, D. W., Mendez-Tellez, P. A.,
Ciesla, N., Desai, S. V., Sevransky, J., Shanholtz, C., Scharfstein, D., Herridge,
M. S., and Pronovost, P. J. (2006). Study Protocol: The Improving
Care of Acute Lung Injury Patients (ICAP) Study. {\it Critical Care}  {\bf 10}, R9.




\bibitem[Qu, Wang, and Wang (2016)]{Qu2016}
Qu, S., Wang, J. L. and  Wang, X. (2016).  Optimal estimation for the functional {C}ox model. {\it The Annals of Statistics}  {\bf 44} 1708--1738.

\bibitem[Ramsay and Dalzell (1991)]{Ramsay1991}
Ramsay, J. O. and Dalzell, C. J. (1991). Some tools for functional data analysis. {\it Journal of the Royal Statistical Society. Series B.	 Methodological}  {\bf 53} 539--572.

\bibitem[Ramsay, Graves, and Hooker (2020)]{fda2020}
Ramsay, J. O., Graves, S. and Hooker, G. (2020). fda: Functional Data Analysis. R package version 5.1.9. https://CRAN.R-project.org/package=fda

\bibitem[Ramsay and Silverman (2005)]{Ramsay2005}
Ramsay, J. O. and Silverman, B. W. (2005). {\it Functional data analysis}. Springer, New York.

\bibitem[Ritov (1990)]{Ritov1990}
 Ritov, Y. (1990). Estimation in a linear regression model with censored data. {\it The Annals of Statistics} {\bf 18} 303--328.


\bibitem[Schumaker (1981)]{Schumaker1981}
Schumaker, L. L. (1981). {\it Spline Functions: Basic Theory}.  Wiley,   New York.

\bibitem[Shen and Wong (1994)]{Shen1994}
Shen, X. and  Wong, W.H. (1994).  Convergence rate of sieve estimates. {\it The Annals of Statistics}  {\bf22}  580--615.

\bibitem[Stone (1982)]{Stone1982}
Stone, C. J. (1982). Optimal global rates of convergence for nonparametric regression. {\it The Annals of Statistics}  {\bf 10}
1040--1053.

\bibitem[Stone (1985)]{Stone1985}
Stone, C. J. (1985). Additive regression and other nonparametric models. {\it The Annals of Statistics}   {\bf 13}  689--705.



\bibitem[Tstats (1990)]{Tstats1990}
Tstatis, A. A. (1990). Estimating regression parameters using linear rank tests for censored data. {\it The Annals of Statistics} {\bf 18}  354--372.

\bibitem[van der Vaart and  Wellner (1996)]{Van1996}
van der Vaart, A. W. and Wellner, J. A. (1996). {\it Weak Convergence and Empirical Processes}.  Springer, New York.

\bibitem[{Vincent et al. (1998)}]{Vincent}
Vincent, J. L., de Mendonça, A., Cantraine, F., Moreno, R., Takala, J., Suter, P. M., Sprung, C. L., Colardyn, F. and Blecher, S. (1998).
Use of the SOFA score to assess the incidence of organ dysfunction/failure in intensive care units: Results of a multicenter, prospective study.
{\it Critical Care Medicine} {\bf 26}, 1793–1800.

\bibitem[Wang, Chiou, and M{\"u}ller (2016)]{Wang2016}
Wang, J. L.,  Chiou, J. M. and M{\"u}ller, H. G. (2016). Functional data analysis. {\it Annual Review of Statistics and Its Application}  {\bf 3} 257--295.

\bibitem[Yang et al. (2020)]{Yang2020}
Yang, S. J., Shin, H., Lee, S. H., Lee, S. and
for the Alzheimer’s Disease Neuroimaging Initiative. (2020).
Functional linear regression model with randomly censored
data: Predicting conversion time to Alzheimer's disease.
{\it Computational Statistics and Data Analysis} {\bf 150} 107009.

\bibitem[Yao and M\"{u}ller (2010)]{Yao2010}
Yao, F. and M\"{u}ller, H. G. (2010). Functional quadratic regression. {\it Biometrika}  {\bf 97} 49--64.

\bibitem[Yao, M\"{u}ller, and  Wang (2005)]{Yao2005}
Yao, F., M\"{u}ller, H. G. and Wang, J. L. (2005).  Functional linear regression analysis for longitudinal data. {\it The Annals of Statistics}  {\bf 33} 2873--2903.

\bibitem[Ying (1993)]{Ying1993}
Ying, Z. (1993). A large sample study of rank estimation for censored regression data. {\it The Annals of Statistics} {\bf 21}  76--99.

\bibitem[Zeng and Lin (2007)]{Zeng2007}
 Zeng, D. and Lin, D. Y. (2007). Efficient estimation for the accelerated failure time model. {\it Journal of the American Statistical Association}  {\bf 102}  1387--1396.


\bibitem[Zhao and Zhang (2017)]{Zhao2017}
Zhao, X. and Zhang, Y. (2017). Asymptotic normality of nonparametric M-estimators with applications to hypothesis testing for panel count data. {\it Statistica Sinica}  {\bf 27} 931--950.

\bibitem[Zhao, Wu, and Yin (2017)]{Zhao20172}
Zhao, X.,  Wu, Y. and Yin, G. (2017). Sieve maximum likelihood estimation for a general class of accelerated hazards models with bundled parameters. {\it Bernoulli} {\bf 23} 3385--3411.

\bibitem[Zhong, Mueller and Wang (2021+)]{Zhong2021}
Zhong, Q., Mueller, J. and Wang, J.-L. (2021).
Deep extended hazard models for survival analysis.
{\it Advances in Neural Information Processing Systems 34 pre-proceedings} (NeurIPS 2021).

\bibitem[Zhong, Mueller and Wang (2021+)]{Zhong2021}
Zhong, Q., Mueller, J. and Wang, J.-L. (2021+).
Deep learning for the partially linear Cox model.
 {\it Annals of Statistics}, in press.









\end{thebibliography}
\end{document}